\documentclass[referee,pdflatex,sn-mathphys-ay]{sn-jnl}% Math and Physical Sciences Author Year Reference Style
\usepackage{graphicx}%
\usepackage{multirow}%
\usepackage{amsmath,amssymb,amsfonts}%
\usepackage{amsthm}%
\usepackage{mathrsfs}%
\usepackage[title]{appendix}%
\usepackage{xcolor}%
\usepackage{textcomp}%
\usepackage{manyfoot}%
\usepackage{booktabs}%
\usepackage{algorithm}%
\usepackage{algorithmicx}%
\usepackage{algpseudocode}%
\usepackage{listings}%

\usepackage{float}
\usepackage{multirow}
\usepackage{subfig}
\usepackage[scaled]{helvet}
\usepackage{rotating}
\usepackage{booktabs}
\usepackage{url}

\newcommand{\argmax}[1]{\underset{#1}{\operatorname{arg}\,\operatorname{max}}\;}

\theoremstyle{thmstyleone}%
\theoremstyle{thmstyletwo}%

\theoremstyle{thmstylethree}%

\begin{document}

\title{Nonparametric modal regression with missing response observations}
%\title[Article Title]{Article Title}


\author*[1,2]{\fnm{Ana} \sur{P\'erez-Gonz\'alez}}\email{anapg@uvigo.gal}

\author[1]{\fnm{Tom\'as R.} \sur{Cotos-Y\'a\~{n}ez}}\email{cotos@uvigo.gal}

\author[2,3]{\fnm{Rosa M.} \sur{Crujeiras}}\email{rosa.crujeiras@usc.es}

\affil*[1]{\orgdiv{Department of Statistics and Operation Research}, \orgname{University of Vigo}, \city{Vigo}, \country{Spain}}

\affil[2]{\orgdiv{Galician Center for Mathematical Research and Technology}, \orgname{CITMAga}, \city{Santiago de Compostela}, \country{Spain}}

\affil[3]{\orgdiv{Department of Statistics, Mathematical Analysis and Optimization}, \orgname{University of Santiago de Compostela}, \city{Santiago de Compostela}, \country{Spain}}

%\affil*[1]{\orgdiv{Department}, \orgname{Organization}, \orgaddress{\street{Street}, \city{City}, \postcode{100190}, \state{State}, \country{Country}}}
%\affil[2]{\orgdiv{Department}, \orgname{Organization}, \orgaddress{\street{Street}, \city{City}, \postcode{10587}, \state{State}, \country{Country}}}
%\affil[3]{\orgdiv{Department}, \orgname{Organization}, \orgaddress{\street{Street}, \city{City}, \postcode{610101}, \state{State}, \country{Country}}}


\abstract{Modal regression has emerged as a flexible alternative to classical regression models when the conditional mean or median are unable to adequately capture the underlying relation between a response and a predictor variable. This approach is particularly useful when the conditional distribution of the response given the covariate presents several modes, so the suitable regression function is a multifunction.  In recent years, some proposals have addressed modal (smooth) regression estimation using kernel methods. In addition, some remarkable extensions to deal with censored, dependent or circular data have been also introduced. However, the case of incomplete samples due to missingness has not been studied in the literature. This paper adapts the nonparametric modal regression tools to handle missing observations in the response, investigating several imputation approaches through an extensive simulation study. The performance in practice of our proposals are also illustrated with two real--data examples.
}




\maketitle
%............................................%
\section{Introduction}
\label{sec:introduction}
%------------------------------------------%
%Par. 1 Why modal regression
The relation between a response variable $Y$ and a covariate $\mathbf X\in\mathbb R^d$ is usually addressed by modelling (parametrically, non parametrically or semiparametrically) the conditional expectation of $Y|\mathbf X$. However, the conditional expectation may not be a suitable parameter to describe the conditional distribution, for instance, when this conditional distribution exhibits a strong asymmetry or a multimodal pattern. In a unimodal but asymmetric situation, quantile regression (considering the median instead of the mean, for instance) may be an alternative, but when more than one mode appears in the conditional distribution, the regression approach should be able to identify the (possibly) several conditional modes.

%Par. 2 First approach for modal regression
Nonparametric conditional mode estimation was early introduced by \cite{scott2015multivariate}, from a kernel density estimation perspective. This approach was used by \cite{Einbek:2006} to propose an adaptation of the mean--shift algorithm, previously presented by \cite{cheng1995} and \cite{comaniciu2002}, for the regression setting. This algorithm maximizes a nonparametric estimator of the conditional density identifying the (regression) branches corresponding to the local conditional modes. \cite{Chen:2016a} derived the asymptotic properties of the estimator provided by the mean--shift algorithm and examined the links of modal regression with other methodologies such as mixture regression or cluster analysis. Subsequently, \cite{Zhou:2019} proposed bandwidth selectors for the modal regression based on bootstrap and cross-validation tools. \cite{alonsopena-2020} presents a comprehensive review on nonparametric multimodal regression with a simulation study comparing the performance of several bandwidth procedures.  For a general overview on modal analysis, see \cite{Chacon2020}.

%Par. 3 Second approach for modal regression
An alternative approach for modal regression is based on structural assumptions for the (unique) conditional mode. For instance, \cite{LEE1989} and \cite{YaoLi2014} consider a linear model for the conditional mode and introduce kernel weights to define an objective function. More complex models, with semiparametric or nonparametric methodologies, are considered by \cite{Yao2012}, \cite{Lv2017}, or  \cite{ULLAH2023}  among others. A drawback of this approach with respect to the nonparametric route is that all these methods are formulated in a unimodal setting, recovering just a single branch of the regression.

%Par 4. Application of modal regression
In addition, modal regression has revealed as a powerful tool for regression analysis in different disciplines such as medicine, where it has been used to investigate conditions such as Alzheimer's disease (see \cite{wang2017regularized}) and the effects of the SARS-CoV-2 virus (see \cite{Ullah2022}). It has also been employed in transport studies (see \cite{Einbek:2006}) and in economic research (see \cite{LEE1989}, \cite{Lee1993}). For a comprehensive overview of the applications of modal estimation, see \cite{Chen2018}.

%Par 5. Motivation of missingness
Although most statistical methods are analyzed and designed to be applied in complete data cases, it is not uncommon to deal with incomplete samples showing missing observations. Missingness may be attributed to a variety of factors, including technical failures in measurement processes or individuals who refuse to answer. This phenomenon is not exclusive to a particular research area and it is usually encountered in fields as diverse as medicine, social studies, and economics. The presence of missing observations may have a profound impact on estimation and inference, justifying the importance of appropriate management strategies to prevent the introduction of biases and ensure the accuracy of statistical analysis.

%Par 6. Literature of missing data (theor + applied)
The statistical literature has analysed the impact of missing data on a wide range of methodologies, including, among others, mean and quantile regression, robust estimation or even goodness-of-fit tests (see \cite{Efromovich2014}; \cite{Chen_quantile}; \cite{boente_estimation}; \cite{Manteiga_Test}). The relevance of this topic is evidenced by the numerous publications dealing with missing data. See, among others, \cite{enders2022applied}, \cite{molenberghs2007missing},  \cite{little2019statistical}, \cite{carpenter2021missing}, \cite{ibrahim2009missing} and \cite{pigott2001review}. These works consider a variety of analytical techniques to deal with the missingness feature. The simplest approach consist in considering just the complete part of the sample (simplified estimation), entailing the deletion of missing observations. An alternative procedure introduces weights that are inversely proportional to the observation probability. This approach is known as Inversely Probability Weighting (IPW). Another possibility is to fill or impute the missing observations using conditional and unconditional imputation methods. A route that has gained relevance in this field is multiple imputation (MI) where the imputation of each missing observation is done via an average estimate of a set of values, see e.g. \cite{Rubin1987}, \cite{van2018flexible} or \cite{Kleinke2020}.

%Par 7. But there is no modal+missing
Although modal regression tools has been adapted to deal with censored data (\cite{khardani2019}), dependent data (\cite{Ullah2022}), measurement error (\cite{zhou2016}, \cite{liHuang2019} or \cite{LiuHuang2024}) and even circular regression models (as illustrated in \cite{alonso-pena_crujeiras_2023}), to the best of our knowledge, the effect of missing data on conditional mode estimation has not yet been studied.

%Par 8. Aim of the work
The performance of the different statistical methodologies is contingent upon the missing data model and the way this feature is considered in the analysis. Thus, the aim of this work is to introduce an adaptation of the nonparametric modal regression approach using the mean--shift algorithm by \cite{Einbek:2006} and \cite{Chen:2016a}, investigating the impact of various missing data methodologies on several modal regression scenarios where the response variable contains missing observations. In addition, an innovative multiple imputation method, specifically tailored for local mode estimation, will be also presented.

%Par 9. Paper organization
This paper is organized follows. Section~\ref{sec:nmr} provides a concise overview of the nonparametric modal regression estimator for complete data. Section~\ref{sec:missing} focuses on different approaches to deal with missingness in the response. Precisely, a simplified estimator, a IPW estimator, and a simple and multiple imputation estimators will be presented. Since our approach uses kernel methods, a bandwidth selector is introduced in Section~\ref{sec:bandwidth}. The numerical performance of the proposed estimators is shown in Section~\ref{sec:simulation}, whereas Section~\ref{sec:data} illustrates their application in two real--data examples. Final comments and some discussion are given in Section~\ref{sec:discussion}. Code and data needed for executing the simulation study and real data examples are available online in Supplementary material Section.

%............................................%
\section{Some background on nonparametric modal regression}
\label{sec:nmr}
Basic ideas on nonparametric modal regression will be introduced in what follows. For a comprehensive review on modal inference, see \cite{Chacon2020}. See also \cite{Einbek:2006}, \cite{Chen:2016a} and \cite{alonsopena-2020} for modal regression using the mean--shift algorithm.

Consider a response variable $Y$ and a covariate $\mathbf X$ with support in $D\subset\mathbb R^d$. Denote by $f(\cdot | \mathbf X=\mathbf x)$ the conditional density of $Y$ for a given $\mathbf X=\mathbf x$. When the aim is to estimate the modes of the conditional density for a given $\mathbf x$, this problem can be viewed as the estimation of a multifunction or a conditional mode set $\mathcal M(\mathbf x)$ defined as
\begin{equation}
	\mathcal{M}\left(\mathbf{x}\right)=\left\lbrace y:\quad\frac{\partial}{\partial y}f\left(y|\mathbf{x}\right)=0, \quad\frac{\partial^2}{\partial^2 y}f\left(y|\mathbf{x}\right)<0\right\rbrace.
\end{equation}
Given that $f\left(y|\mathbf{x}\right)=\frac{f\left(\mathbf{x},y\right)}{f\left(\mathbf{x}\right)}$, the set $\mathcal M\left(\mathbf{x}\right)$ can be expressed as:

\begin{equation}
	\mathcal M\left(\mathbf{x}\right)=\left\lbrace y:\quad\frac{\partial}{\partial y}f\left(\mathbf{x},y\right)=0,\quad \frac{\partial^2}{\partial^2 y}f\left(\mathbf{x},y\right)<0\right\rbrace.
\end{equation}

The nonparametric estimation of $\mathcal M(\mathbf x)$ can be approached by considering a kernel density estimator of $f\left(\mathbf{x},y\right)$, labelled as $\widehat f_n\left(\mathbf{x},y\right)$ (see \cite{scott2015multivariate}) which provides a plug-in estimator of $\mathcal M\left(\mathbf{x}\right)$:
 \begin{equation}
	\widehat{\mathcal{M}}\left(\mathbf{x}\right)=\left\lbrace y:\quad\frac{\partial}{\partial y}\widehat f_n\left(\mathbf{x},y\right)=0, \quad\frac{\partial^2}{\partial^2 y}\widehat f_n\left(\mathbf{x},y\right)<0\right\rbrace,
\end{equation}
with 
 \begin{equation}
	\widehat{f}_n\left(\mathbf{x},y\right)=\frac{1}{nh_1^dh_2}\sum_{i=1}^{n} K_1\left(\frac{\|\mathbf{x}-\mathbf{X}_i\|}{h_{1}}\right)K_2\left(\frac{y-Y_i}{h_{2}}\right),
 \label{eq:kde}
\end{equation}
being $\|\cdot\|$ the $L_2$-norm in $\mathbb R^d$, $K_1$ and $K_2$ symmetric smooth kernel functions, $h_1$ and $h_2$ are bandwidth parameters acting on the response and on the covariate observations, and $\{\left(\mathbf{X}_i,Y_i\right)_{i=1}^n\}$ denotes the sample observations. Usually, both $K_1$ and $K_2$ are taken as Gaussian kernels, since the selection of the smoothing functions does not show a relevant impact on the estimators. Regarding the bandwidth parameters, although further discussion will be given in Section 4, it is worth noticing the distinct role that both $h_1$ and $h_2$ present in modal regression. On the one hand, $h_1$, the bandwidth acting on the covariate observations, presents the usual oversmoothing (undersmoothing) behaviour for large (respectively, low) values over all the possible branches of the regression. On the other hand, $h_2$ is acting on the response observations, and controls the number of branches: large values of $h_2$ lead to a single branch (as a limit case) whereas too small values of $h_2$ provide an extremely fragmented pattern (see \cite{alonsopena-2020} for an illustration with complete data samples).

Although the statement of the problem seems simple, the local maxima (set of local modes) of the conditional density estimation cannot (in general) be obtained analytically and numerical methods are required for computing $\widehat{\mathcal M}\left(\mathbf{x}\right)$. In this setting, \cite{Einbek:2006} proposed an algorithm based on the mean--shift algorithm to estimate the set of conditional modes. The conditional mean--shift algorithm for complete data, suggested by \cite{Einbek:2006}, is described below:

%-----------------------------------------------%
\bigskip
\hrule
\textbf{Algorithm 0}: Conditional mean--shift\\
\hrule
\begin{enumerate}
%\item[]\textbf{Data}: sample $\{\left(\mathbf{X}_i,Y_i\right)\}_{i=1}^n$. Bandwidth parameters $\left(h_1,h_2\right)$ and the kernel functions $K_1$ and $K_2$.
		\item[Step 1.] Initialize mesh points $M=\{\mathbf x_j,j=1,\ldots,m\} \subseteq D\subset R^{d}$
		\item[Step 2.] For each $\mathbf{x}\in M$, select starting points $\{y_0^{1}\left(\mathbf{x}\right),y_0^{2}\left(\mathbf{x}\right),...,y_0^{p}\left(\mathbf{x}\right)\}$,
			\item[Step 3.]  For a fixed $\mathbf x$ and for each $(\mathbf{x},y_0^{k}\left(\mathbf{x}\right))$, with $k=1,...,p$,  initialize $t=0$ and update $y_{t+1}^{k}$ using the following iteration procedure:
			\begin{equation} \nonumber
				y_{t+1}^{k}\left(\mathbf{x}\right)=\frac{\sum_{i=1}^{n} Y_iK_1\left(\frac{\|\mathbf{x}-\mathbf{X}_i\|}{h_{1}}\right)K_2\left(\frac{y_t^{k}\left(\mathbf{x}\right)-Y_i}{h_{2}}\right)}{\sum_{i=1}^{n} K_1\left(\frac{\|\mathbf{x}-\mathbf{X}_i\|}{h_{1}}\right)K_2\left(\frac{y_t^k\left(\mathbf{x}\right)-Y_i}{h_{2}}\right)}, \quad \textrm{with}\quad t=0,1,...
			\end{equation}
   Repeat Step 3 until the convergence criterion is reached at 
   $\{y^{1}\left(\mathbf{x}\right),y^{2}\left(\mathbf{x}\right),\cdots,y^{p}\left(\mathbf{x}\right)\}$.
\end{enumerate}
\hrule
%\bigskip
%-----------------------------------------%

The modal regression multifunction estimator at point $\mathbf{x}$ provided by Algorithm 0 is given by
\[
\widehat{\mathcal M}\left(x\right)=\{y^{1}\left(\mathbf{x}\right),y^{2}\left(\mathbf{x}\right),\cdots,y^{p}\left(\mathbf{x}\right)\}
\]
This estimator, which is a random set, was further studied by \cite{Chen:2016a}, considering Gaussian kernels and, for simplicity, the same bandwidth parameter for the response and the covariate ($h_1=h_2$), to obtain asymptotic error bounds and construct confidence intervals. The asymptotic analysis carried out by \cite{Chen:2016a} is based on an error criterion measuring the approximation of $\widehat{\mathcal M}\left(x\right)$ to $\mathcal M\left(x\right)$, using the Hausdorff distance between two sets, defined as:
\begin{equation}
	\mbox{Haus}\left(A,B\right)= \inf_{r\geq 0}\left\{r: A \subseteq B\oplus r, B \subseteq A\oplus r\right\},
\end{equation}
with $A\oplus r=\left\{\mathbf{x}: d(\mathbf{x},A)\leq r\right\}$, and $d(\mathbf{x},A)=\inf_{\mathbf{y}\in A} \left\|\mathbf{x}-\mathbf{y} \right\|$.

Note that the use of a kernel density estimator for the function $\widehat f_n\left(\mathbf{x},y\right)$ in Algorithm 0 requires the selection of (at least) a bandwidth parameter. \cite{Einbek:2006},  \cite{Chen:2016a}  and  \cite{Zhou:2019} address this choice from different methodological perspectives, including cross--validation, uniform prediction sets and residual squares criterion. A numerical comparison of the different proposals was provided by \cite{alonsopena-2020}.

%---------------------------------------------------%
\section{Modal regression with missing response data}
\label{sec:missing}
%---------------------------------------------------%
In order to deal with the possible missing responses in our data sample $\{\mathbf X_i,Y_i\}_{i=1}^n$, a new variable $\delta$ (indicating whether $Y$ is observed or not) should be introduced. Specifically, $\delta_i=1$ if $Y_i$ is observed and $\delta_i=0$ otherwise. We consider a missing at random  (MAR) model for $\delta$ (see \cite{little2019statistical}), given by
\begin{equation}	\mathbb P(\delta=1|\mathbf{X},Y)=\mathbb P(\delta=1|\mathbf{X})=p\left(\mathbf{X}\right)
	\label{missingmodel}
\end{equation}
Some issues should be hilighted regarding the modelling of the missingness mechanism. Specifically, \cite{Bojinov} observed that this mechanism can be disregarded in the context of Bayesian or direct--likelihood inference when the MAR assumption holds. Nevertheless, the frequentist approach is capable of ignoring the missingness mechanism under a more robust condition of missing completely at random (MCAR). For missing data, likelihood--based frequentist inference is only considered valid when the missingness mechanism generates datasets satisfying the MAR condition. This is in accordance with the findings of \cite{little2019statistical}, \cite{little2021missing} and \cite{Seaman2013}.
%As pointed \cite{Bojinov}, the missing mechanism can be ignored for Bayesian or direct-likelihood inference under the MAR  assumption. However the frequentist approach can do it under a stronger condition of  Missing Completely at random (MCAR). Only likelihood-based frequentist inference would be valid when missing mechanism generates missing at random datasets, \cite{little2019statistical}, \cite{little2021missing}, \cite{Seaman2013}.

 The performance of the techniques employed in the analysis of incomplete datasets are contingent upon the methodology employed and the nature of the missing data. As it has been mentioned in the Introduction, some of these techniques entail the deletion of a portion of the sample, which may be done in a listwise, casewise, or pairwise manner. While their implementation is relatively straightforward, the inference process requires strong assumptions and may result in a significant reduction in the number of records. As an alternative, IPW methods are derived from sampling statistics, assigning non--response weights, which are inversely proportional to the probability of observation (see (\ref{missingmodel})). These quantities must be estimated from the sample with any of the existing procedures: logistic or nonparametric regression, among others (see \cite{Seaman2011}, \cite{bianco2011asymptotic}, \cite{Shao2016}).

An alternative approach to address the non--response is to impute the missing observations.  If the missing observation is replaced by a single value, the technique is referred to as single imputation. Some mechanisms employ values derived from other observations (\emph{hot--deck} imputation), mean substitution, or regression--based imputation (see \cite{little2019statistical}, \cite{enders2022applied} and \cite{molenberghs2007missing}). The use of auxiliary variables may enhance the effectiveness of the imputation process, particularly when the predictions are in close alignment with the actual data. However, this approach may potentially result in an underestimation of the variance associated with the imputed data (see \cite{van2018flexible}, \cite{little2019statistical}). In order to address the potential issue of low dispersion of imputed observations, multiple imputation was devised. This approach involves the repeated simulation of values for the missing data in order to obtain (at least) two complete data sets. Each of the imputed datasets is subjected to analysis using standard statistical techniques. The results obtained are then combined to create a single estimator, from which inferences can be drawn. Multiple imputation is outlined in detail in \cite{van2018flexible}.
A recent paper  of \cite{Little2024} shows, from a non-technical perspective, the strengths and weaknesses of above-mentioned methods.

In what follows, specific adaptations of the conditional mean--shift algorithm (namely Algorithm 0) will be presented for simplified, IPW, simple imputation and a suitably adapted multiple imputation approaches.

%-----------------------------------------%
\subsection{Simplified estimator}
The simplified estimation can be obtained considering only the complete observations. In this setting, the kernel density estimation of $f\left(\mathbf{x},y\right)$ is given by:
\begin{equation}
	\widehat{f}_{S,n}\left(\mathbf{x},y\right)=\frac{1}{h_1^dh_2\sum_{i=1}^{n}\delta_i}\sum_{i=1}^{n}\delta_i K_1\left(\frac{\|\mathbf{x}-\mathbf{X}_i\|}{h_{1}}\right)K_2\left(\frac{y-Y_i}{h_{2}}\right)
 \label{denscc}
\end{equation}

Following similar calculations as in the complete case (see  \cite{zhou2016} or \cite{alonso-pena_crujeiras_2023}  among others) with Gaussian kernels $K_1$ and $K_2$,  the local maximum of  the complete--case estimation density, $ \widehat{f}_{S,n} $, must satisfy:

%\begin{equation*}
%	\nabla\widehat{f}_{S,n}\left(\mathbf{x},y\right)=\frac{2c_{K_2}}{h_1^dh_2^2\sum_{i=1}^{n}\delta_i}\sum_{i=1}^{n}\delta_i %K_1\left(\frac{\|\mathbf{x}-\mathbf{X}_i\|}{h_{1}}\right)K_2\left(\frac{y-Y_i}{h_{2}}\right) \left[ %\frac{\sum_{i=1}^{n}\delta_i K_1\left(\frac{\|\mathbf{x}-\mathbf{X}_i\|}{h_{1}}\right)K_2\left(\frac{y-Y_i}{h_{2}}\right)Y_i}%{\sum_{i=1}^{n}\delta_i K_1\left(\frac{\|\mathbf{x}-\mathbf{X}_i\|}{h_{1}}\right)K_2\left(\frac{y-Y_i}{h_{2}}\right)}-%y\right]
% \label{denscc}
%\end{equation*}

%\begin{equation*}
%
%	\nabla\widehat{f}_{S,n}\left(\mathbf{x},y\right)=\frac{2c_{K_2}}{h_2} \widehat{f}_{S,n}\left(\mathbf{x},y\right)\left[ \frac{\sum_{i=1}^{n}\delta_i K_1\left(\frac{\|\mathbf{x}-\mathbf{X}_i\|}{h_{1}}\right)K_2\left(\frac{y-Y_i}{h_{2}}\right)Y_i}{\sum_{i=1}^{n}\delta_i K_1\left(\frac{\|\mathbf{x}-\mathbf{X}_i\|}{h_{1}}\right)K_2\left(\frac{y-Y_i}{h_{2}}\right)}-y\right]
 %\label{denscc2}
%\end{equation*}

\begin{equation*}
	\quad\frac{\partial}{\partial  y}\widehat{f}_{S,n}\left(\mathbf{x},y\right)=\frac{1}{h_2} \widehat{f}_{S,n}\left(\mathbf{x},y\right)\left[ \frac{\sum_{i=1}^{n}\delta_i K_1\left(\frac{\|\mathbf{x}-\mathbf{X}_i\|}{h_{1}}\right)K_2\left(\frac{y-Y_i}{h_{2}}\right)Y_i}{\sum_{i=1}^{n}\delta_i K_1\left(\frac{\|\mathbf{x}-\mathbf{X}_i\|}{h_{1}}\right)K_2\left(\frac{y-Y_i}{h_{2}}\right)}-y\right]=0
	%\label{denscc2}
\end{equation*}

Then, the mean shift vector is proportional to the normalized density gradient estimate and the algorithm can be adapted just modifying Step 3 as follows:

\bigskip
\hrule
\textbf{Algorithm 1}: Simplified  mean--shift
\hrule
\begin{enumerate}
%    \item Initialize mesh points $M \subseteq R^{d}$
%    \item For each $\mathbf{x}\in M$, select starting points $\{y_0^{1}\left(\mathbf{x}\right),y_0^{2}\left(\mathbf{x}\right),...,y_0^{p}\left(\mathbf{x}\right)\}$,
    \item[Step 3.]  For a fixed $\mathbf x$ and for each $(\mathbf{x},y_0^{k}\left(\mathbf{x}\right))$, with $k=1,...,p$,  initialize $t=0$ and update $y_{t+1}^{k}$ using the following iteration procedure:
        \begin{equation} \nonumber
		y^k_{t+1}\left(\mathbf{x}\right)=\frac{\sum_{i=1}^{n}\delta_i Y_iK_1\left(\frac{\|\mathbf{x}-\mathbf{X}_i\|}{h_{1}}\right)K_2\left(\frac{y_t^{k}\left(\mathbf{x}\right)-Y_i}{h_{2}}\right)}{\sum_{i=1}^{n} \delta_i K_1\left(\frac{\|\mathbf{x}-\mathbf{X}_i\|}{h_{1}}\right)K_2\left(\frac{y_t^k\left(\mathbf{x}\right)-Y_i}{h_{2}}\right)}, \quad \textrm{with}\quad t=0,1,...
	\end{equation}
     Repeat Step 3 until the convergence criterion is reached at
      $\{y^{1}\left(\mathbf{x}\right),y^{2}\left(\mathbf{x}\right),\cdots,y^{p}\left(\mathbf{x}\right)\}$
\end{enumerate}
\hrule
\bigskip

The kernel density estimator based on the completed sample (\ref{denscc}) may be biased if the missing data are not MCAR (see \cite{Dubnicka-2009}). For this reason, other alternatives must be considered.   

%----------------------------------------------------------%
\subsection{Inversely Probability Weighting estimator}
 An alternative to the simplified estimator that is widely used in the context of  missing data is the IPW method, following the idea of  \cite{HorvitzThompson1952}, where each observation is inversely weighted by the corresponding propensity score (\ref{missingmodel}). Hence, the nonparametric kernel density estimator of the joint density is given by:

\begin{equation}  \nonumber
	\widehat{f}_{IPW,n}\left(\mathbf{x},y\right)=\frac{1}{h_1^dh_2\sum_{i=1}^{n}\frac{\delta_i}{p\left(\mathbf{X}_i\right)}}\sum_{i=1}^{n}\frac{\delta_i}{p\left(\mathbf{X}_i\right)} K_1\left(\frac{\|\mathbf{x}-\mathbf{X}_i\|}{h_{1}}\right)K_2\left(\frac{y-Y_i}{h_{2}}\right)
 % \label{densipw}
\end{equation}

The theoretical properties of the IPW estimator  of the density function have been studied by \cite{Dubnicka-2009} in the univariate case, although the estimator can be extended to the multivariate case. Analogously to the previous case, the gradient of IPW density estimation can be obtained and the mean--shift algorithm is adapted as follows:

%--------------------------------------------%
\bigskip
\hrule
\textbf{Algorithm 2}: IPW conditional mean--shift
\hrule
\begin{enumerate}
%    \item Initialize mesh points $M \subseteq R^{d}$
%    \item For each $\mathbf{x}\in M$, select starting points $\{y_0^{1}\left(\mathbf{x}\right),y_0^{2}\left(\mathbf{x}\right),...,y_0^{p}\left(\mathbf{x}\right)\},$
    \item[Step 3.]  For a fixed $\mathbf x$ and for each $(\mathbf{x},y_0^{k}\left(\mathbf{x}\right))$, with $k=1,...,p$,  initialize $t=0$ and update $y_{t+1}^{k}$ using the following iteration procedure:
	\begin{equation}  \nonumber
		y^{t+1}\left(\mathbf{x}\right)=\frac{\sum_{i=1}^{n}\frac{\delta_i}{p\left(\mathbf{X}_i\right)} Y_i K_1\left(\frac{\|\mathbf{x}-\mathbf{X}_i\|}{h_{1}}\right)K_2\left(\frac{y_t^{k}\left(\mathbf{x}\right)-Y_i}{h_{2}}\right)}{\sum_{i=1}^{n} \frac{\delta_i}{p\left(\mathbf{X}_i\right)} K_1\left(\frac{\|\mathbf{x}-\mathbf{X}_i\|}{h_{1}}\right)K_2\left(\frac{y_t^k\left(\mathbf{x}\right)-Y_i}{h_{2}}\right)}, \quad \textrm{with}\quad t=0,1,...
	\end{equation}
 Repeat Step 3 until the convergence criterion is reached at
      $\{y^{1}\left(\mathbf{x}\right),y^{2}\left(\mathbf{x}\right),\cdots,y^{p}\left(\mathbf{x}\right)\}$
\end{enumerate}
\hrule
\bigskip

The missing probability, $p\left(\mathbf{X}_i\right)$, is usually unknown and it must estimated consistently. Such an estimator can be obtained parametrically, for example,  using  a  logistic regression or nonparametrically with a kernel estimator, $\widehat p(\mathbf X_i)$.  It is worth noticing a counter--intuitive phenomenon that has been observed with applying the IPW estimator with $\widehat p(\mathbf X_i)$: the nonparametric estimation of the missing probability reduces the variance of IPW estimators even when the probability is known (for further details, see references \cite{bianco2011asymptotic}, \cite{robins1995analysis} or \cite{wang1998local} among others).

%-----------------------------------------------%
\subsection{Imputed simple estimator}
Imputation methods are another route to deal with missing observations. The conventional approach is based on mean imputation, whereby the conditional or unconditional mean of the variable containing missing values is estimated and employed to replace the missing values. In the context of modal regression, this approach may not be adequate, particularly when the conditional distribution of the response variable is asymmetric, resulting in a significant discrepancy between the conditional mean and mode. Accordingly, an imputation method based on the simplified estimator (Algorithm 1) is deemed an appropriate means of filling the non--observed data. This estimator is constructed in two stages: first, Algorithm 1 is employed to forecast the mode of the response variable for $\delta=0$. This process results in the generation of a complete sample, which can then be subjected to the conditional mean--shift algorithm for complete data (Algorithm 0).

%------------------------------------%
\bigskip
\hrule
\textbf{Algorithm 3}: Simple Imputation conditional mean--shift
\hrule
\begin{enumerate}
%	\item [] Data: Sample $\{\left(\mathbf{X}_i,\delta_i,Y_i\right)\}_{i=1}^n$. Bandwidth parameters $\left(h_1,h_2\right)h$ and the kernel function is chosen as Gaussian.
    \item[Step 1:] Imputation of non--observed data
	\begin{enumerate} 
    	\item[1] For each $Y_i$ with $\delta_i=0$, consider the predictor variable $\mathbf{X}_i$ and select starting points
            $\{y_{i,0}^{1},y_{i,0}^{2},\cdots,y_{i,0}^{p}\}$,
		\item[2] Apply Algorithm 1 with the complete sample and the previous starting points
		\begin{equation} \nonumber
    	y_{i,t+1}^k\left(\mathbf{X}_i\right)=\frac{\sum_{j=1}^{n}\delta_j Y_j K_1\left(\frac{\|\mathbf{X}_i-\mathbf{X}_j\|}{h_{1}}\right)K_2\left(\frac{y_{i,t}^{k}-Y_j}{h_{2}}\right)}{\sum_{i=1}^{n} \delta_i K_1\left(\frac{\|\mathbf{x}-\mathbf{X}_i\|}{h_{1}}\right)K_2\left(\frac{y_{i,t}^k-Y_i}{h_{2}}\right)}, \quad \textrm{with}\quad t=0,1,...
    	\end{equation}
	   \item[3] Take $\widehat{\mathcal M}(\mathbf{X}_i)=\left\{\widehat{Y}_{i,j}\right\}_{j=1}^{k_i}$ the set of $k_i$ estimated modes in $\mathbf{X}_i$.
          \item[4] Evaluate the conditional density $\widehat{f}_n\left(\cdot|\mathbf{X}_i\right)$ at the elements of $\widehat{\mathcal M}(X_i)$ and obtain $\widehat{Y}_{i,j_0}$ with $$j_0=\argmax{j} \left\{\widehat{f}_n\left(\widehat{Y}_{i,j}|\mathbf{X}_i\right): 1\leq j \leq k_i   \right\}$$
	The imputed observations are completed in the following way:
	\begin{equation*}
	   \widetilde{Y}_i=\delta_iY_i+(1-\delta_i)\widehat{Y}_{i,j_0}
	\end{equation*}
\end{enumerate}
\item[Step 2:] Conditional mode estimation
\item[] Once the sample is completed,  $\left\{ \left(\mathbf{X}_i,\widetilde{Y}_i\right)\right\}_{i=1}^n$ we apply the mean--shift algorithm for complete data to this sample, that is, Algorithm 1 with $\delta_i=1$, for $i=1,\ldots,n$. 
\end{enumerate}
\hrule
\bigskip

In Step 2, note that the mean--shift vector is proportional to the density of the imputed sample, represented by the set of points $ \left\{ \left(\mathbf{X}_i, \widetilde{Y}_i\right)\right\}_{i=1}^n$. 

 In general, the imputed data based in a single value (mean, regression, a random value, etc) does not represent all the variability of the original data.  Consequently, a potential limitation of the simple imputation method is that standard errors may be underestimated (see \cite{van2018flexible} or \cite{Rubin1987}, among others). However, multiple imputation allows for handling uncertainty associated with potential imputations. In particular, the density estimator based on the mean-imputed sample  does not present a satisfactory asymptotic behavior, given that the imputed sample is treated as if it was a complete data set. \cite{Titterington} compared the mean-integrated squared error  for different estimators of the density function with missing at random data, and the usefulness of the  multiple imputation was highlighted in the paper. 

%-------------------------------------------------%
\subsection{Multiple imputation estimator} 
A novel multiple imputation method for nonparametric modal regression is presented in this section.  In general, the pooling step of multiple imputation algorithms consist of averaging estimates. Nevertheless, in a modal context, averaging is not the adequate strategy. 
%In our case, however, this combination method is not appropriate because we are estimating conditional modes.

The proposed multiple imputation estimator considers two phases. The first step is to apply Algorithm 1 to the available data $\{\left(\mathbf{X}_i,Y_i,\delta_i=1\right)\}_{i=1}^n$, where the mesh points considered are the sample points $\mathbf{X}_i$ where $Y_i$ is missing and $Y-$ coordinates are equispaced. So, a modal set is estimated for each sample point $\mathbf{X}_i$ where $\delta_i=0$. From these modal sets, a number $B$ of values are randomly drawn to fill the missing responses for each incomplete observation ($\delta_i=0$), and thus $B$ different completed (by imputation) samples are obtained $\left\{\left\{\left(\mathbf{X}_i,\widetilde{Y}_{i}^j\right)\right\}_{i=1}^n\right\}_{j=1}^B$. 

In the next step, Algorithm 0 is applied to these $B$ filled samples and $B$ modal sets are estimated in each new mesh point set. In this case, the algorithm for selecting the mesh points is the same as for the complete data. The estimated modal sets can be different because the $B$ input data sets are different.  
Finally, for each mesh point, the algorithm combines these $B$ modal sets using procedures based on techniques to determine the number of unconditional modes as in \cite{AmeijeirasAlonso:2021} and a new estimated modal set is computed.

\bigskip
\hrule
\textbf{Algorithm 4}: Multiple imputation conditional mean--shift 
\hrule
\begin{enumerate}
 % \item [] Data:Sample$ \{\left(\mathbf{X}_i,\delta_i,Y_i\right)\}_{i=1}^n$. The kernel function, $K$, is chosen as Gaussian. \newline
    \item [] Repeat Step 1 $B$ times.
    \item[Step 1:] Imputation of the sample
	\begin{enumerate} 
    	\item[1.1] For each index $i$ where $Y_i$ is missing, consider $\mathbf{X}_i$ and  select the  starting points. 
		 $\{y_{i,0}^{1},y_{i,0}^{2},\cdots,y_{i,0}^{p}\}$, as the same way from previous algorithms but with the value of the covariate fixed at $\mathbf{X}_i$  with $\delta_i=0$.
		\item[1.2] Apply Algorithm 1 with the complete sample and the previous starting points
		\begin{equation}
			y_{i,t+1}^k=\frac{\sum_{j=1}^{n}\delta_j Y_j K_1\left(\frac{\|\mathbf{X}_i-\mathbf{X}_j\|}{h_{1}}\right)K_2\left(\frac{y_{i,t}^{k}-Y_j}{h_{2}}\right)}{\sum_{j=1}^{n}  \delta_j K_1\left(\frac{\|\mathbf{X}_i-\mathbf{X}_j\|}{h_{1}} \right) K_2\left(\frac{y_{i,t}^k-Y_i}{h_{2}}\right)}
		\end{equation}
		\item[1.3] Let $\widehat{\mathcal M}(\mathbf{X}_i)=\left\{\widehat{Y}_{i,j}\right\}_{j=1}^{k_i}$ be the set of $k_i$ estimated modes in $\mathbf{X}_i$.
		\item[1.4] Compute the conditional density $\widehat{f}_n\left(\cdot/\mathbf{X}_i\right)$  at points of $\widehat{M}(\mathbf{X}_i)$ and now select randomly one value with weights proportionals to $\left(\widehat{f}_n\left(\widehat{Y}_{i,1}/\mathbf{X}_i\right), \cdots, \widehat{f}_n\left(\widehat{Y}_{i,k_i}/\mathbf{X}_i\right)\right)$, say $\widetilde{Y}_{i}$.   
	\end{enumerate}
	\item[Step 2:] Estimation with the complete samples\newline
 Once the sample is completed,  $\left\{\left\{\left(\mathbf{X}_i,\widetilde{Y}_{i}^j\right)\right\}_{i=1}^n\right\}_{j=1}^B$ with $\widetilde{Y}_i^j=\delta_iY_i+(1-\delta_i)\widehat{Y}_{i}$, we apply the mean--shift algorithm (Algorithm 0) for complete data to these samples.
		
\item[Step 3:] Combine the results\newline
For each mesh point $\mathbf{x_j}\in M$, from the previous step, we have obtained $B$ sets of  conditional modes from the previous $B$ imputations and it is necessary to merge them to obtain a estimated modal set for each one. For this objective, we join the $B$ sets and we use unconditional strategies to compute the number of modes and the identification of them. In particular,  a procedure which combines the smoothing and the excess mass, proposed by \cite{AmeijeirasAlonso:2021} is used.
\end{enumerate}
\hrule
\bigskip

%-----------------------------------------%
\section{Bandwidth selection}
\label{sec:bandwidth}

The selection of the bandwidth parameters $h_1$ and $h_2$ in (\ref{eq:kde}) is crucial for the implementation of kernel methods. In particular, two approaches have been considered for the conditional modal regression: the use of a bandwidth selector for kernel density estimation and the use of a specific bandwidth that takes into account the nature of multimodal sets. An extensive simulation study on multimodal bandwidth selectors in the complete data scenario has been carried out by \cite{alonsopena-2020}.
%The key point in all kernel methods is the selection of the bandwidth parameters. In particular, for the conditional modal regression, two point of view have been addressed: use bandwidth selector for the kernel density estimation or an specific bandwidth taking into account the nature of the multimode sets. 

As observed by \cite{Einbek:2006}, the bandwidth selectors based on the conditional density estimation are prone to oversmoothing. A corrected method based on a hybrid bandwidth selection rule introduced by \cite{bashtannyk2001bandwidth} can be used instead.
%In the paper of \cite{Einbek:2006} noted the bandwidth selectors based on the conditional density estimation are oversmoothed. A corrected method based on a hybrid bandwidth selection rule of \cite{bashtannyk2001bandwidth} is used in the example.

\cite{Chen:2016a} proposed a bandwidth selector based on the volume of the predictions sets. Given a level $\alpha$, the prediction set is defined as:
$$\mathcal {P}_{1-\alpha}=\{\left(\mathbf x,y\right),(\mathbf x\in D,y\in \mathcal M\left(\mathbf x\right)\oplus\epsilon_{1-\alpha}\},$$
where
$$\epsilon_{1-\alpha}=\inf\{\epsilon\geq0:\mathbb P\left(d\left(Y,\mathcal M(\mathbf x)\right)>\epsilon\right)\leq \alpha\}.$$
The authors assumed  $h_1=h_2=h>0$ and the univariate bandwidth selector of $h$ is obtained minimizing the volume (given by the Lebesge measure) of a kernel density estimation of $\mathcal P_{1-\alpha}$ from:
    \begin{equation}
    \arg \min_{h\geq 0} \mbox{Vol}\left(\widehat{\mathcal P}_{1-\alpha,h}\right)=\widehat \epsilon_{1-\alpha,h}\int_{\mathbf x\in D}\widehat N_h\left(\mathbf x\right)d\mathbf x,
    \end{equation}
where $\widehat N_h\left(\mathbf x\right)$ is the estimated number of local modes at $\mathbf X=\mathbf x$. This selector has two limitations: the parameter $\alpha$ must be selected and the assumption $h_1=h_2$ may be very restrictive because, as pointed out in the introduction and by  \cite{alonsopena-2020}, the role of the parameters is different, one controlling the smoothing and other the number of modes. 

From the computational point of view, \cite{Zhou:2019} conducted a comparative study of several bandwidth selectors for nonparametric modal regression with one-dimensional covariate and propose two alternatives based on the cross-validation criterion and other which minimized an empirical version of the Integrated Squared Error (ISE) based on the Hausdorff distance between the estimated and true modal set. As the true modal set is not available, Bootstrap sampling was used to estimate the ISE as follows:

\begin{equation}
 \widehat{ISE}\left(h_1,h_2\right)=\frac{1}{R}\sum_{r=1}^R\frac{1}{n}\sum_{i=1}^{n}\mbox{Hauss}\{\hat M_{h_1,h_2}^r\left(\mathbf x\right),\hat M^*\left(\mathbf x\right)\},
\end{equation}

where $\hat M_{h_1,h_2}^r$ is the estimator based on the $r$th bootstrap sample and $M^*$ is the estimator based on a finite mixture model. 

Later, \cite{alonsopena-2020} compared these selectors and concluded that the CV criterion adapted to modal regression performs well in many situations, but when the number of modes varies with $\mathbf{X}$ may be the estimator is not able to adequately capture the modal structure.

Actually, multimodal estimation with a d-dimensional covariate finds a problem in bandwidth selection. The existing proposals assume a univariate covariate as \cite{Zhou:2019} or a univariate kernel evaluated in $L_2-$ norm in $\mathbb R^d$ to avoid the multidimensional smoothing parameter selection. 

%From the computational point of view, \cite{Zhou:2019} conducted a comparative study of bandwidth selectors for nonparametric modal regression. Moreover, they proposed two methods based in a cross-validation and  Bootstrap to select  bandwidths. The authors proposed two selectors based on cross validation (CV) method and the  minimization of the empirical version of the Integrated Squared Error of the Hausdorff distance between $\widehat M\left(x\right)$ and $M\left(x\right)$. 

In the missing response context, the bandwidth selector should be also adapted to adequately handle the missingness. Based on the cross--validation proposal by \cite{Zhou:2019}, an IPW version can be formulated as follows: consider $(h_1,h_2)$ the values that minimize the cross--validation function:
\begin{equation}
    \label{CV.mod}
    CV\left(h_1,h_2\right)=n^{-1}\sum_{i=1}^{n}\delta_id^2\left(\widehat{M}_{-i}\left(\mathbf{X}_i\right),Y_i\right)N^2_{-i}\left(\mathbf{X}_i\right)\frac{w\left(\mathbf{X}_i\right)}{p\left(\mathbf{X}_i\right)}.
\end{equation}
where $d\left(A,b\right)$ denotes the Hausdorff distance between a set $A$ and a  point $b$ as the minimum Euclidean distance between $b$ y $a \in A$. The term $N\left(\mathbf{X}_i\right)$ denotes the size of $\widehat{\mathcal M}\left(\mathbf{X}_i\right)$. Note that the IPW version of the cross--validation criterion considers both the indicator of observed responses, $\delta_i$ and the missing probability $p(\mathbf X_i)$. As commented in section 3.2, the  missing probability must be estimated and different methodologies can be used. Parametric assumptions, as logistic model, or a nonparametric regression of the $\delta$ variable over $X$, among other alternatives, can be used.

%based on CV criterion and which takes to account the size of the mode test. The reasons for this choice are twofold: its lower computational cost and the fact that it is less stable than the first method proposed. For complete data, the selector is obtained from:

%\begin{equation}
%	CV\left(h_1,h_2\right)=n^{-1}\sum_{i=1}^{n}d^2\left(\widehat{M}_{-i}\left(\mathbf{X}_i\right),Y_i\right)N^2_{-i}\left(\mathbf{X}_i\right)w\left(\mathbf{X}_i\right)
%\end{equation}

%where $d\left(A,b\right)$ denotes the distance between a set $A$ and a  point $b$ as the minimum Euclidean distance between $b$ y $a \in A$. The term $N\left(\mathbf{X}_i\right)$ denotes the size of $\widehat{M}\left(\mathbf{X}_i\right)$.

%In our case, we consider a IPW version of cross validation criterion:
%\begin{equation}
%    \label{CV.mod}
%    CV\left(h_1,h_2\right)=n^{-1}\sum_{i=1}^{n}\delta_id^2\left(\widehat{M}_{-i}\left(\mathbf{X}_i\right),Y_i\right)N^2_{-i}\left(\mathbf{X}_i\right)\frac{w\left(\mathbf{X}_i\right)}{p\left(\mathbf{X}_i\right)}.
%\end{equation}

 %............................................%
\section{Simulation experiments}
\label{sec:simulation}
A simulation study has been carried out in order to study the performance of the modal regression nonparametric estimators with missing responses introduced in Section~\ref{sec:missing}. Numerical studies where conducted in \cite{R.software}.

%This section presents the findings of a simulation study that compares the performance of nonparametric estimators of multimodal regression, as defined in Section 3, under different models. The numerical studies were conducted using the \cite{R.software}. 
%This section presents the findings of a simulation study to compare, under different models, the performance of the nonparametric estimators of the multimodal regression defined in Section 3.The \cite{R.software} software is selected to carry out all the numerical studies. 

\subsection{Simulation scenarios}
In all cases, we generate independent and identically distributed data from the model:

$$Y_i= 2 \sin(2\pi X_i) + \epsilon^{k,a}_i, \ 1\leq i \leq n$$
where $X_i \sim U(0,1)$ and $\epsilon^{k,a}$ is the error variable following a mixed Gaussian distribution. More precisely, $\epsilon^{k,a}\sim kZ_1(a) + (1-k)Z_2$ with $Z_1(a) \sim N(\mu_1+(\mu_2-\mu_1)a, \sigma_1)$, $Z_2 \sim N(\mu_2,\sigma_2)$ and $k \in [0,1]$ and $a \in [-1,1]$. The remaining parameters are $\mu_1=-1.5$, $\mu_2=1.5$ and $\sigma_1=\sigma_2=0.5$ and $n=200$.

The parameter $k$ serves to regulate the ratio of the model density $k/(1-k)$ (and therefore of the modes). The values of $a$ determine the distance between the Gaussian means of the mixture ($a=0$, representing a separation of three units and $a=1$ indicating no separation).

\begin{figure}[ht]
    \centering
    \includegraphics[width=0.4\textwidth]{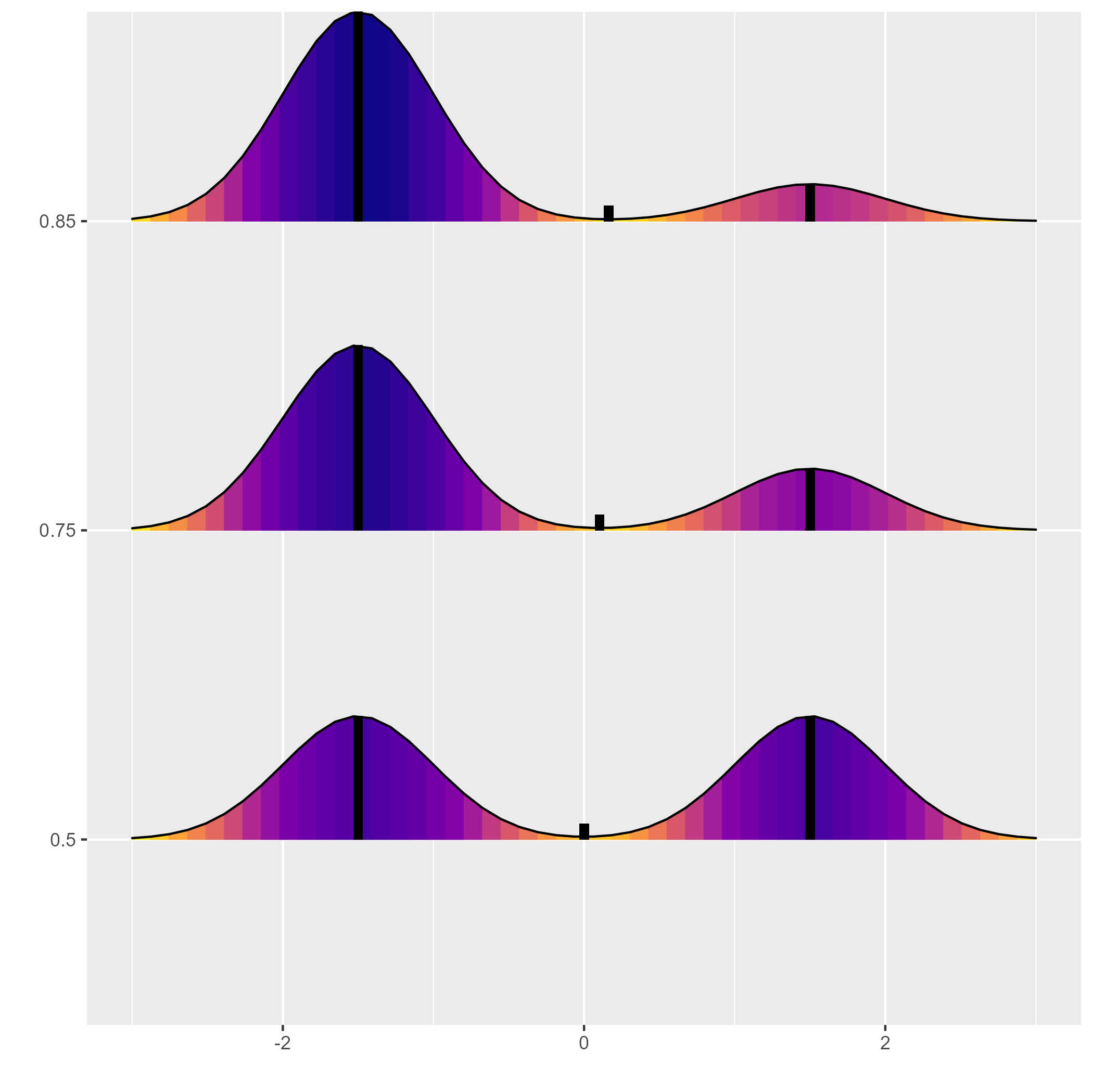}
    \includegraphics[width=0.4\textwidth]{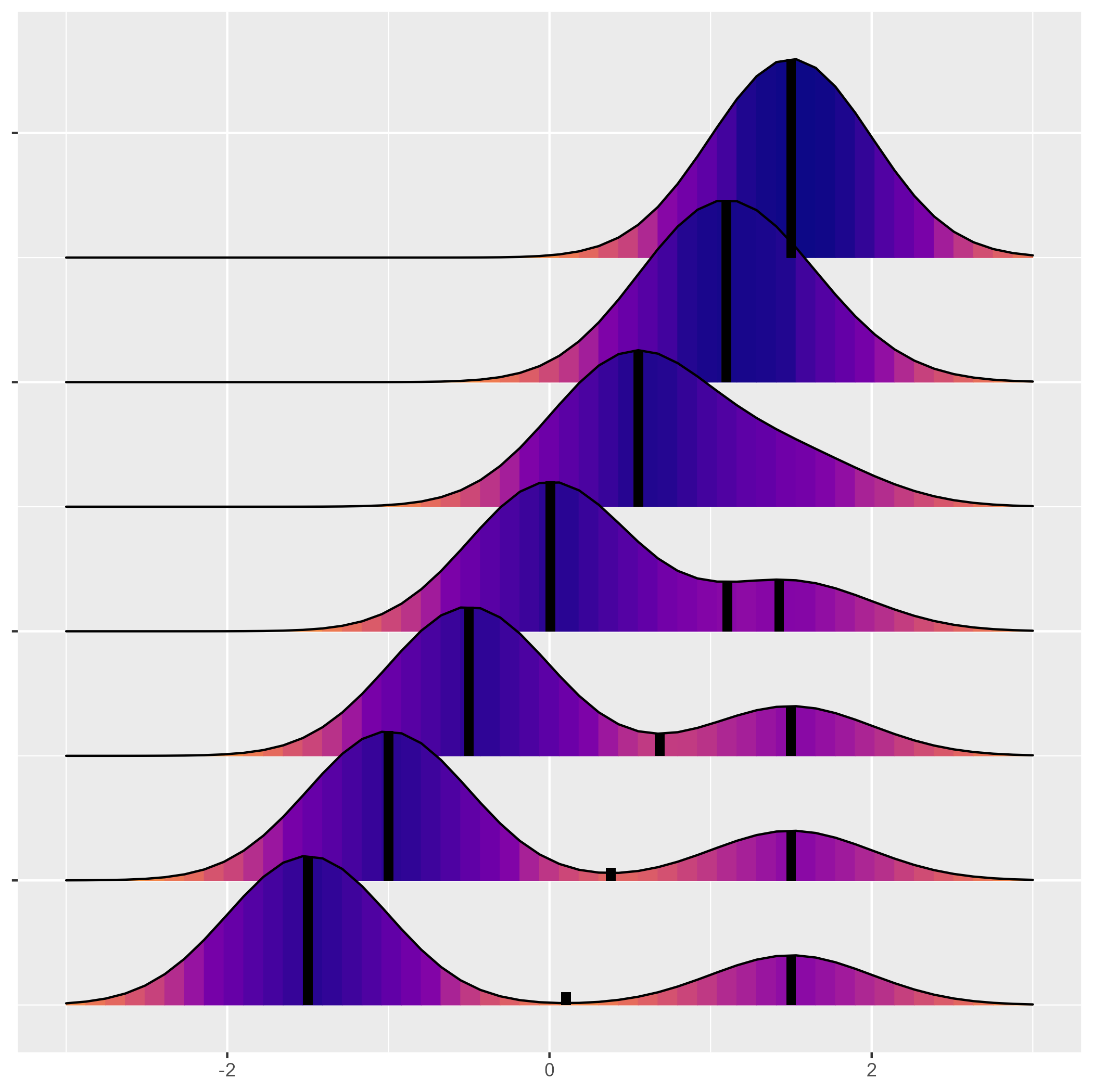}
    \caption{Mixture of Gaussian distributions for scenario 1 (left) and scenario 2 (right). For scenario 1, from top to bottom: $k=0.85,0.75,0.5$. For scenario 2, $k=0.75$ and from top to bottom, $a \in \{1, 5/6, 4/6, 0.5, 2/6, 1/6, 0\}$. Vertical black lines show the local modes.}   
    \label{figura:ESC1ESC2}
\end{figure}

\begin{figure}[ht]
\centering
    \includegraphics[width=0.32\textwidth]{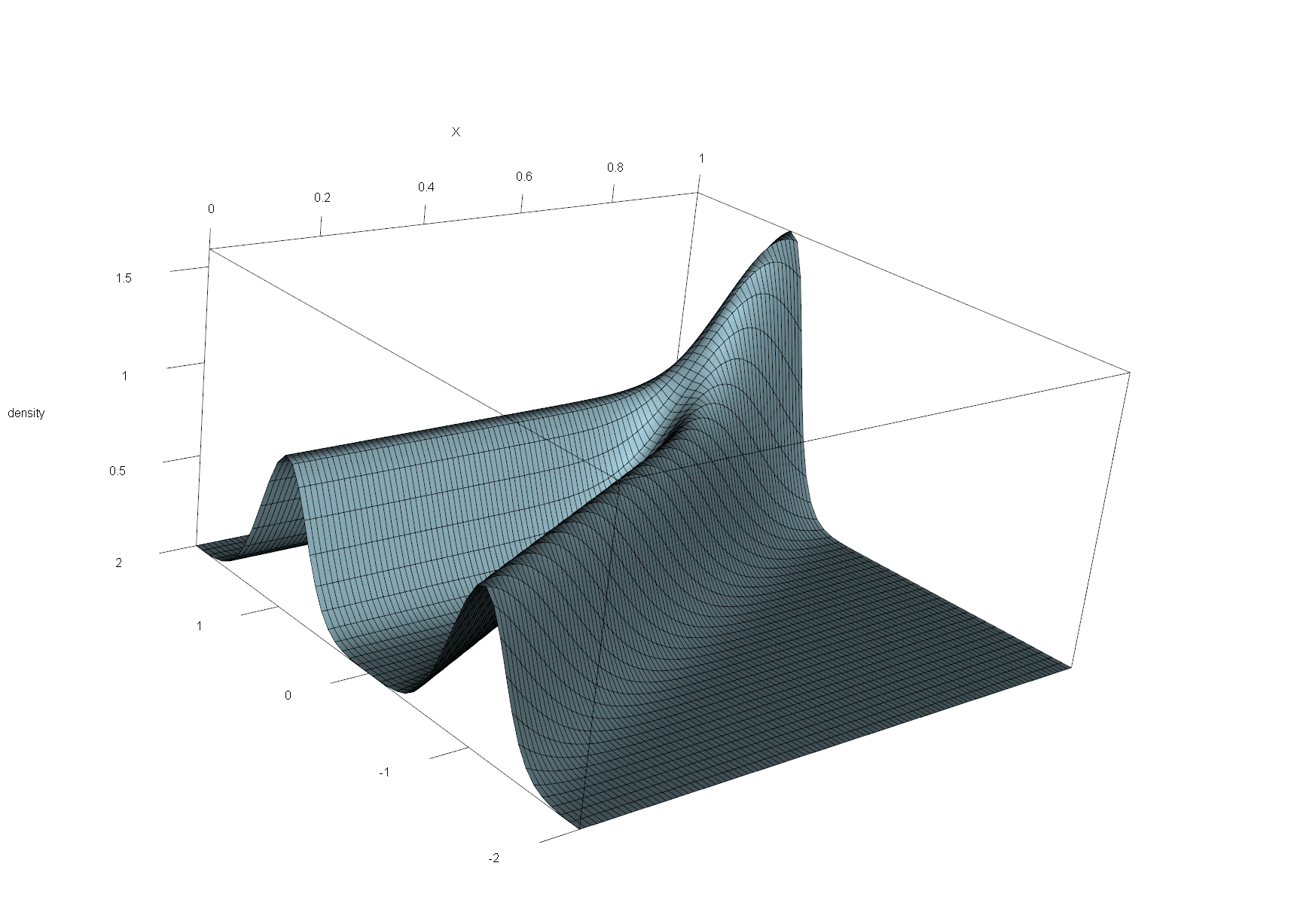}
    \includegraphics[width=0.32\textwidth]{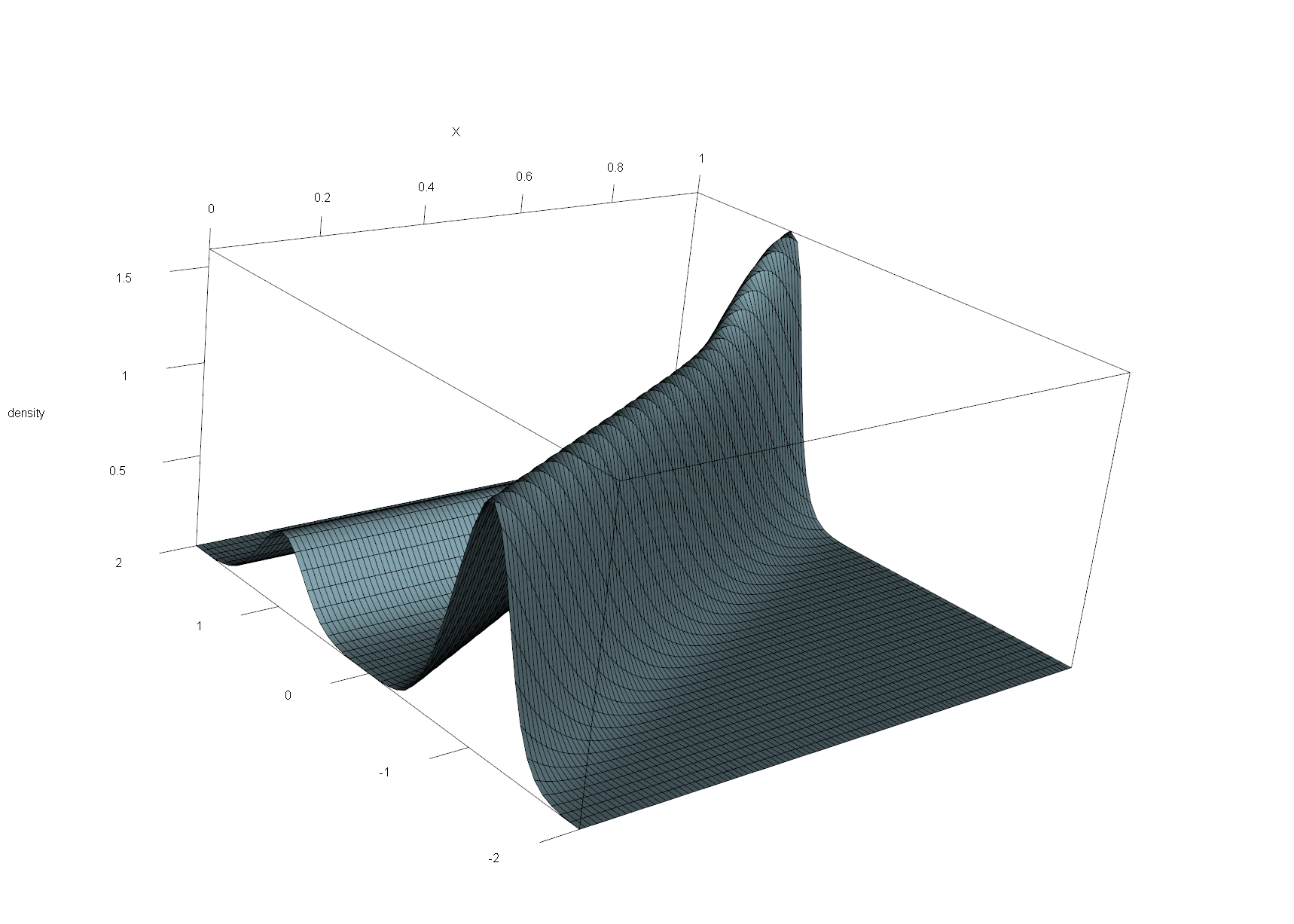}
    \includegraphics[width=0.32\textwidth]{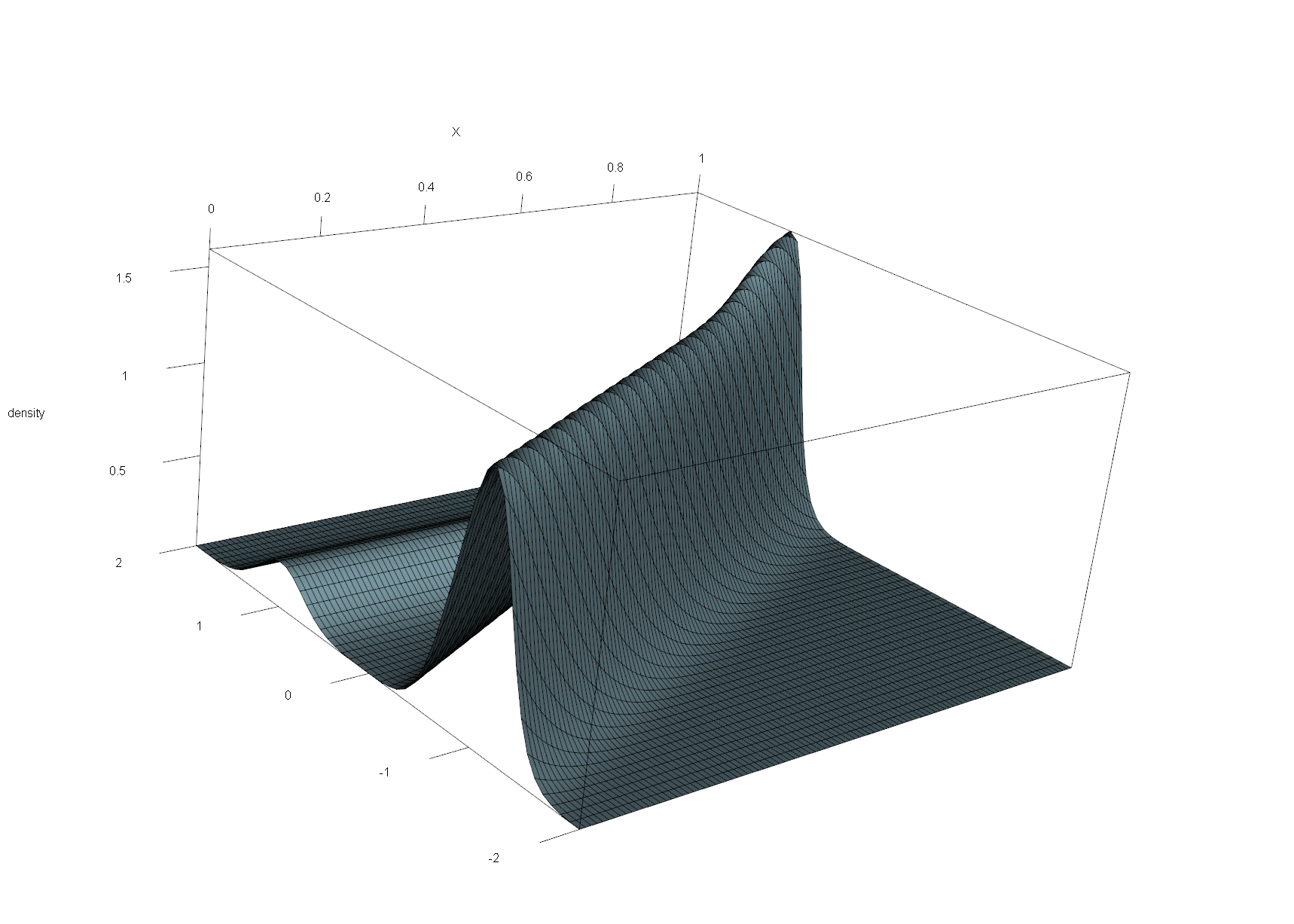}
    \caption {Mixture of Gaussian distributions in scenario 3. From left to right, $k=0.5,0.75$ and $k=0.85$.}
    \label{figura:ESC3}
\end{figure}

\begin{figure}[ht]
\centering
    \includegraphics[width=5cm]{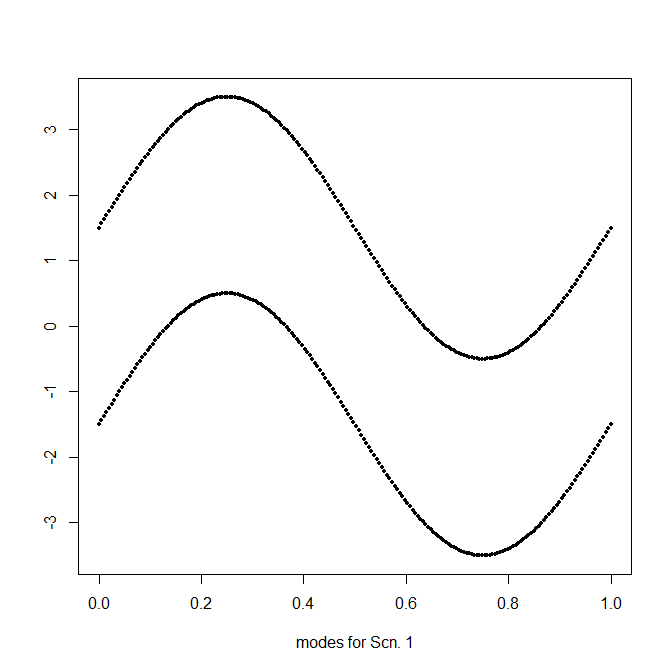}
    \includegraphics[width=5cm]{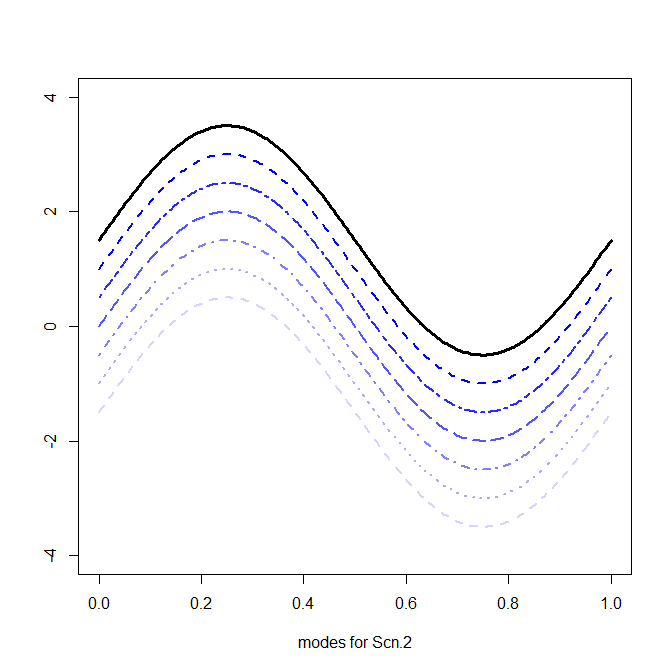}
    \includegraphics[width=5cm]{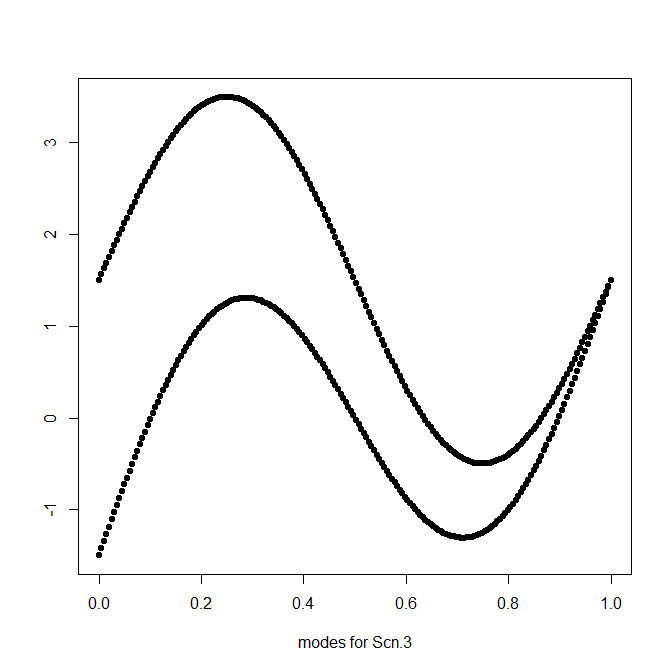}
    \caption {Modal regression curves (branches of the modal regression) for scenarios 1 to 3, from left to right.}
        \label{figura:modes}
\end{figure}

The following scenarios are considered:
\begin{enumerate}
	\item[Scn.1] $\epsilon^{k,a}=\epsilon^{k,0}$, i.e., a mixture of two Gaussian distributions (Figure \ref{figura:ESC1ESC2}), with two fixed modes at $\mu_1,\mu_2$ for $k=0.5,0.75\text{ and }0.85$ (Figure \ref{figura:modes}) with high density values (depending on $k$) and a third mode with values at $0, 0.10 \text{ and } 0.16$ depending on $k$ and with density values close to zero.  
	\item[Scn.2] $\epsilon^{k,a}=\epsilon^{0.75,a}$, i.e., a mixture of two Gaussian distributions (Figure \ref{figura:ESC1ESC2}), with modes depending of $a=0,\frac{1}{6},\frac{2}{6},\frac{3}{6},\frac{4}{6},\frac{5}{6}\text{ and }1$ (Figure \ref{figura:modes}).
	\item[Scn.3] $\epsilon^{k,a}=\epsilon^{\left(k,X_i\right)}, \ 1\leq i \leq n$, i.e., a mixture of two Gaussian distributions (Figure \ref{figura:ESC3}) whose main modes vary as $\{\mu_1+(\mu_2-\mu_1)X_i,\mu_2\}$ and $k=0.5,0.75\text{ and }0.85$ (Figure \ref{figura:modes}).
\end{enumerate}

Figures \ref{figura:ESC1ESC2} and \ref{figura:ESC3} illustrate the densities of the mixtures for scenarios 1 to 3. The values of $a$ and $k$ are employed to simulate multimodal patterns and to examine the impact of the transition from unimodality to multimodality on the various estimators. It should be noted that the error distribution of scenario 3 (Figure \ref{figura:ESC3} ) depends on the predictor variable $X$ and the number of local modes varies with $X$. This is the most complex scenario, even with complete data. 

The location of the conditional modes is markedly distinct under these models, as illustrated in Figure \ref{figura:modes}. In scenarios 1 and 2, the conditional modes are shown to be parallel with a fixed distance for scenario 1. In scenario 2, the distance between the conditional modes decreases. In contrast, scenario 3 shows how the conditional modes become closer depending on the predictor variable values. 

Missing data (\ref{missingmodel}) are generated according four model configurations:

\begin{enumerate}
	\item[M1.] $p\left(x\right)=0.6+0.3\cos(\pi x)$.%1/(1+exp(-2*(x-0.3)^2))$.
	\item[M2.] $p\left(x\right)=0.6+0.3\cos(2 \pi x)$.%1/(1+exp(2*(x-0.8)^2))$.
	\item[M3.] $p\left(x\right)=0.7+0.3\cos(2\pi x^2)$.
	\item[M4.] $p\left(x\right)=0.75$.%0.6+0.3\cos(2*\pi*x)$.
\end{enumerate}

\begin{figure}[ht]
\centering
\includegraphics[width=7.5cm]{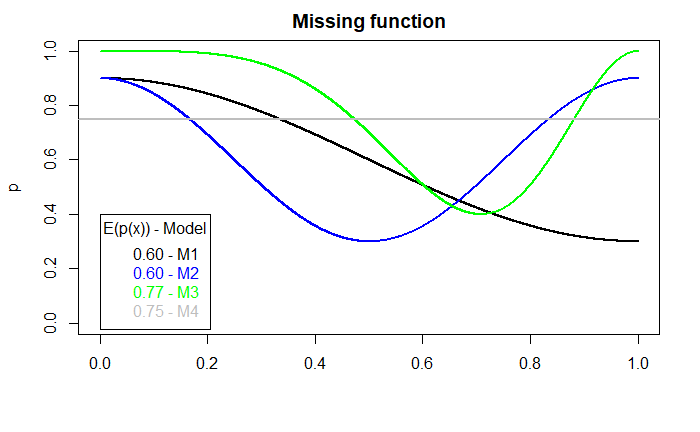}
\caption {Missing function models M1 to M4 (black, blue, green and grey, respectively).}
\label{figura:MissingModel}
\end{figure}

Figure \ref{figura:MissingModel} shows the probability of to be missing for M1 to M4 models. M4 model assumes a MCAR pattern with (constant) 25\% of missing data. M1 to M3 models show different situations under a MAR scheme. The percentage of observed data is similar in all situations with values between 60\% and 75\%. 

In all the examples, bandwidth selection has been performed with the IPW cross--validation criterion introduced in expression 
 (\ref{CV.mod}).

% Densidades so escenarios

%The bandwidth selector was by IPW version of cross validation criterion (\ref{CV.mod}). %cross-validation \cite{Zhou:2019}.

A measure of the global error is computed based on the average of the squared Hausdorff distance between the set of theoretical and estimated modes over a grid in the interval $[0,1]$ of length $m=200$, given by:
%We compute a measure of the global error based on an average the squared Hausdorff distance between the set of theoretical and estimated modes over a grid in $[0,1]$ of length 200, given by:
$$\mbox{ASE}\left(\widehat{\mathcal M}\right)=\frac{1}{B}\sum_{b=1}^B {\sum_{j=1}^{m}{\mbox{Hauss}\left( \widehat{\mathcal M}({x}_j),\mathcal M({x}_j)\right)^2}}\Delta$$
where $\Delta$ is the partition resolution of the grid ($\Delta=1/200$) and $B$ is the number of samples. In all cases, $B=500$.

The simulations, to observe the effect of missingness, are also carried out assuming that the complete sample is observed ($p\left(x\right)=1\text{ }\forall x$), as outlined in Algorithm 0, with the estimator labeled as $\textbf{C}$. The estimators obtained by Algorithms 1 and 2, simplified and inversely probability-weighted,  are designated as $\textbf{S}$ and $\textbf{W}$, respectively. The imputed versions of Algorithms 3 and 4, namely simple and multiple imputation, are labeled $\textbf{SI}$ and $\textbf{MI}$, respectively.
% We perform the simulations in the case where the complete sample is observed, Algorithm 0, the estimator is labelled with $\textbf{C}$. The estimators obtained by Algorithm 1 an 2, Simplified and Inversely Probability weighted, are named as $\textbf{S}$ and $\textbf{W}$ respectively. The imputed versions of Algorithm 3 and 4, simple and multiple imputation, are labelled as $\textbf{SI}$ and $\textbf{MI}$.

There are different methods to select the starting points: whole sample, equidistant, random, proportional to quantiles, etc. For the imputation methods ($\textbf{SI}$ and $\textbf{MI}$) we consider the complete subsample as mesh points for the first step. For the next steps of these estimators and the other algorithms ( $\textbf{C}$, $\textbf{S}$ and $\textbf{W}$) a fixed number of starting points equally spaced was considered; as pointed  \cite{Einbek:2006}, large enough to reach the modes even more than once. 

%-------------------------------------------%
\subsection{Results for scenario 1}
\begin{figure}[H]
\centering
\includegraphics[width=0.9\linewidth]{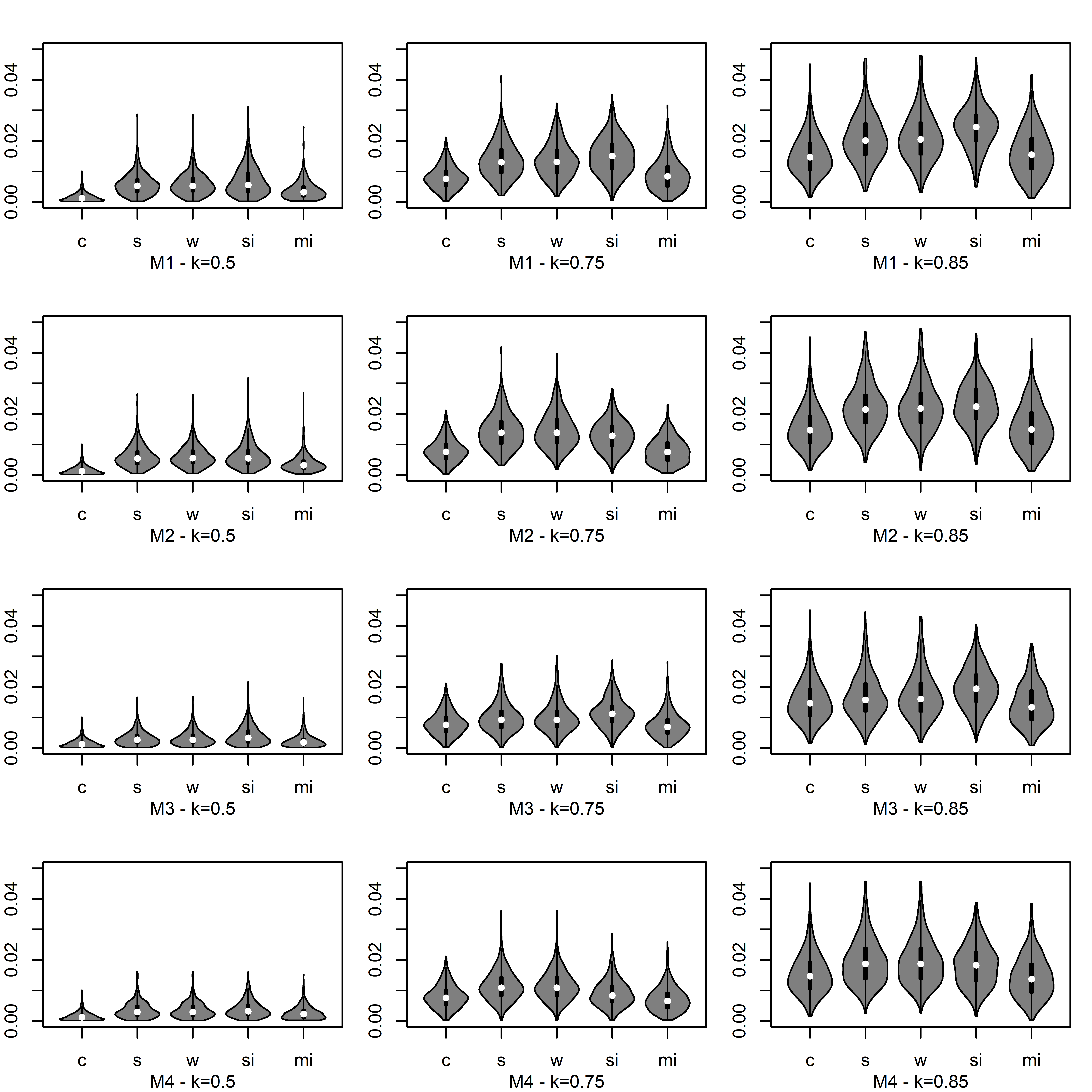}
\caption{Vioplot of $\mbox{ASE}(\mathcal M)$ for scenario 1 and  missing models M1-M4, with $k=0.5,0.75,0.85$.}
\label{figura:ESC1-vioplot}
\end{figure}

In Scenario 1, the modes can clearly distinguished for $k=0.5$. In this case, the behaviour of the simplified estimator $\textbf{S}$, the IPW estimator $\textbf{W}$ and the simplified imputation estimator $\textbf{SI}$ is similar for all missing data models (M1 to M4). However, the multiple imputation estimator $\textbf{MI}$ estimator performs better than its competitors.  The results for the $\mbox{ASE}\left(\widehat{\mathcal M}\right)$ are presented in Table~\ref{tabla:ESC1}. For $k=0.75$ and $k=0.85$, as illustrated on the left side of Figure \ref{figura:ESC1ESC2}, the distribution is almost unimodal and $\textbf{S}$ and $\textbf{W}$ estimators exhibit similar performance. In contrast, the $\textbf{SI}$ estimator demonstrates the poorest performance, while the \textbf{MI} estimator exhibits the greatest efficacy. The vioplot for squared error of Hausdorff distance are showed in Figure~\ref{figura:ESC1-vioplot}. For $k=0.75$ and $k=0.85$, where the presence of two modes introduces additional complexity, the $\textbf{MI}$ estimator demonstrates superior performance compared to the other estimators. For $k=0.5$, all the proposed solutions are found to be similar. 

\subsection{Results for scenario 2}
Results for Scenario 2 are presented in Table~\ref{tabla:ESC2}. Multiple imputation \textbf{MI} presents again superior performance for the missing data models. In certain instances, the $\textbf{MI}$ method appears to outperform complete data, however this is an artifact of oversmoothing as previously mentioned in the Introduction. As anticipated, there is a relevant influence of the $a$ parameter on the comparative evaluation of the algorithms. For large values of $a$, yielding a close to unimodality setting, the $\mbox{ASE}\left(\widehat{\mathcal M}\right)$ values are relative small, yet the behaviour of the proposed estimators remains similar. Nevertheless, for relatively small values of $a$, with a pronounced bimodality, the distinction is more clear. In this case, there is a discernible enhancement in the performance of the \textbf{MI} approach, particularly for $a=1/6$.

\begin{figure}[H]
\centering
\includegraphics[width=0.9\linewidth]{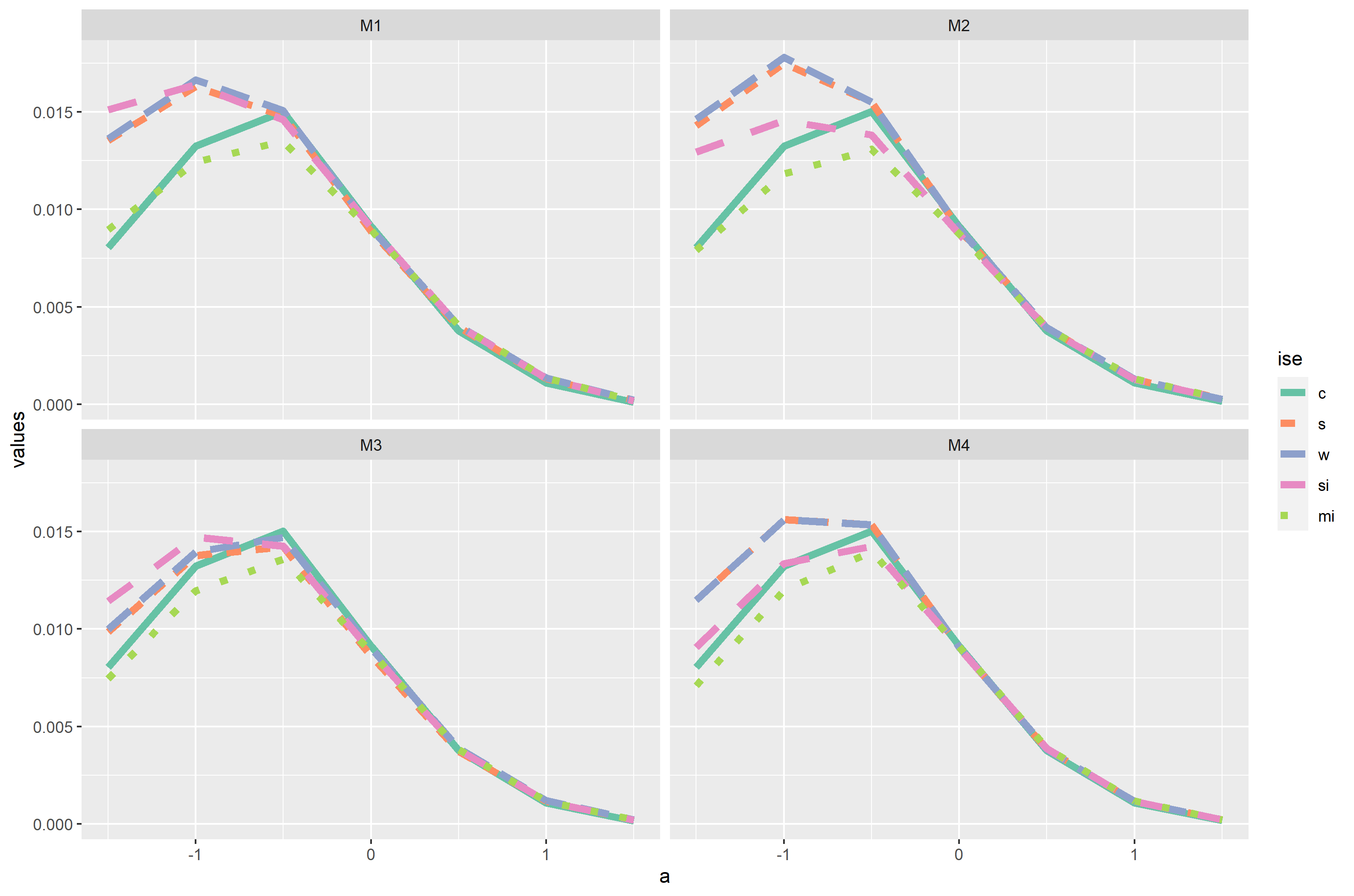}
\caption {$\mbox{ASE}\left(\widehat{\mathcal M}\right)$ for missing models M1 to M4, and several values of $a$ in $[-0.5,0.5]$ in scenario 2.}
\label{figura:ESC2-ise} 
\end{figure}

%With respect to Scenario 2, see Table \ref{tabla:ESC2}, the multiple imputation results are again the best for the missing data models. In some cases, the $\textbf{MI}$ even appear to be better than complete data, but this is an effect of over-smoothing.
%The effect of $a$ parameter is, as our would expect, determining for the comparative of the algorithms. For large values of $a$, close to uni-modality, the $ASE\left(\widehat{M}\right)$ values are small but the behaviour of  proposed estimators is similar. However, for small values of $a$, with a marked bi-modality, the difference is more significant; with a clear improvement of \textbf{MI} for the case $a=1/6$.
Figure~\ref{figura:ESC2-ise} illustrates the behaviour of the $\mbox{ASE}\left(\widehat{\mathcal M}\right)$ with respect to a grid values of $a$ in $\{0,1/6,\ldots,1\}$. For all the missing data models M1 to M4, the errors are larger when the distribution is multimodal, and the values of $a$ less than zero, the $\mbox{ASE}\left(\widehat{\mathcal M}\right)$ increases as the parameter rises towards zero. In this situations, the estimators are unable to distinguish between a multimodal or unimodal conditional distribution. Once the parameter $a$ exceeds zero, the estimators can better capture the shape and the values of $\mbox{ASE}\left(\widehat{\mathcal M}\right)$ decrease drastically. In all cases, the $\mbox{ASE}\left(\widehat{\mathcal M}\right)$ reports values for \textbf{MI} that are lower than those of the other estimators, with a more pronounced discrepancy observed for small values of $a$.

\subsection{Results for scenario 3}
\begin{figure}[H]
\centering
\includegraphics[width=0.9\linewidth]{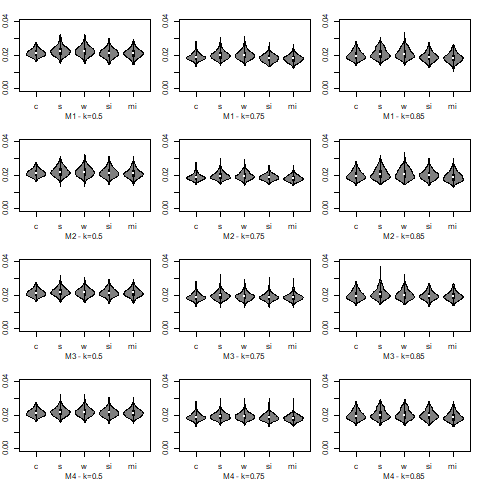}
\caption{Vioplot of $\mbox{ASE}(\mathcal M)$ for scenario 3 and  missing models M1-M4, with $k=0.5,0.75,0.85$.}
\label{figura:ESC3-vioplot}
\end{figure}

The most complex conditional distribution is the one simulated under Scenario 3. The values of $\mbox{ASE}\left(\widehat{\mathcal M}\right)$ in Table~\ref{tabla:ESC3} are  notably high in all cases, particularly for large values of $k$ where the presence of multiple modes is more pronounced. While the differences are less notable, the $\textbf{MI}$ estimator emerges as the best option among the proposed estimators for all missing models. Figure~\ref{figura:ESC3-vioplot} shows the behavior of the error through vioplots with a slight improvement of $\textbf{MI}$ estimator across all examined situations

\begin{sidewaystable}%[H]
% \centering
%\resizebox{17cm}{!} {
\begin{tabular}{rrrrrrrrrrrrrrrr}
\toprule
	 & \multicolumn{5}{c}{\pmb{$k= 0.5$}} & \multicolumn{5}{c}{\pmb{$k= 0.75$}} & \multicolumn{5}{c}{\pmb{$k=0.85$}}\\
\pmb{Model} & \textbf{C} & \textbf{S} & \textbf{W} & \textbf{SI} & \textbf{MI} & \textbf{C} & \textbf{S} & \textbf{W} & \textbf{SI} & \textbf{MI} & \textbf{C} & \textbf{S} & \textbf{W} & \textbf{SI} & \textbf{MI} \\
\midrule
% & c & s & w & si & mi & c & s & w & si & mi & c & s & w & si & mi \\ 
%  \hline
  \textbf{M1} & 1.6 & 5.7 & 5.8 & 6.9 & 3.8 & 8.0 & 13.5 & 13.6 & 15.1 & 9.0 & 15.5 & 20.8 & 21.1 & 24.4 & 16.1 \\ 
  \textbf{M2} & 1.6 & 6.0 & 6.1 & 6.4 & 3.7 & 8.0 & 14.3 & 14.6 & 12.9 & 7.9 & 15.5 & 22.2 & 22.3 & 23.0 & 15.7 \\ 
  \textbf{M3} & 1.6 & 3.1 & 3.2 & 4.1 & 2.3 & 8.0 & 9.9 & 10.0 & 11.4 & 7.4 & 15.5 & 16.9 & 17.3 & 19.7 & 14.2 \\ 
  \textbf{M4} & 1.6 & 3.5 & 3.5 & 3.9 & 2.7 & 8.0 & 11.5 & 11.5 & 9.1 & 7.1 & 15.5 & 19.4 & 19.4 & 18.3 & 14.4 \\ 
\bottomrule
\end{tabular}
%}
\caption{$\mbox{ASE}\left(\widehat{\mathcal M}\right)$ ($\times 1000$) for complete (\textbf{C}), simplified (\textbf{S}), IPW (\textbf{W}), single imputation (\textbf{SI}) and multiple imputation (\textbf{MI}) methods in scenario 1.} 
\label{tabla:ESC1}
\end{sidewaystable}

%\subsection{Scenery 2}
% latex table generated in R 4.1.1 by xtable 1.8-4 package
% Fri Mar 25 13:38:19 2022
\begin{sidewaystable}%[H]
%\resizebox{17cm}{!} {
\begin{tabular}{rrrrrrrrrrrrrrrr}
\toprule
	 & \multicolumn{5}{c}{\pmb{$a= \frac{1}{6}$}} & \multicolumn{5}{c}{\pmb{$a= \frac{3}{6}$}} & \multicolumn{5}{c}{\pmb{$a=\frac{5}{6}$}}\\
\pmb{Model} & \textbf{C} & \textbf{S} & \textbf{W} & \textbf{SI} & \textbf{MI} & \textbf{C} & \textbf{S} & \textbf{W} & \textbf{SI} & \textbf{MI} & \textbf{C} & \textbf{S} & \textbf{W} & \textbf{SI} & \textbf{MI} \\
\midrule
% & c & s & w & si & mi & c & s & w & si & mi & c & s & w & si & mi \\ 
%  \hline
  \textbf{M1} & 13.2 & 16.3 & 16.7 & 16.4 & 12.4 & 9.1 & 8.9 & 9.0 & 9.1 & 8.9 & 1.1 & 1.3 & 1.4 & 1.3 & 1.4 \\ 
  \textbf{M2} & 13.2 & 17.5 & 17.8 & 14.6 & 11.8 & 9.1 & 9.0 & 9.0 & 8.7 & 8.8 & 1.1 & 1.3 & 1.3 & 1.3 & 1.3 \\ 
  \textbf{M3} & 13.2 & 13.8 & 14.0 & 14.7 & 12.0 & 9.1 & 8.6 & 9.0 & 8.8 & 8.8 & 1.1 & 1.2 & 1.2 & 1.2 & 1.2 \\ 
  \textbf{M4} & 13.2 & 15.6 & 15.6 & 13.4 & 12.0 & 9.1 & 9.1 & 9.1 & 9.0 & 9.1 & 1.1 & 1.2 & 1.2 & 1.2 & 1.2 \\
   \hline
\end{tabular}
%}
\caption{$\mbox{ASE}\left(\widehat{\mathcal M}\right)$ ($\times 1000$) for complete (\textbf{C}), simplified (\textbf{S}), IPW (\textbf{W}), single imputation (\textbf{SI}) and multiple imputation (\textbf{MI}) methods in scenario 2 and $a=\frac{1}{6},\frac{3}{6},\frac{5}{6}$.} 
\label{tabla:ESC2}
\end{sidewaystable}

% \textcolor{blue}{Falta grafico con todos os casos de a1 --> de -1.5 a 1.5, paso =0.5. Non hai vioplot son 28 graficos}

%\subsection{Scenery 3}
\begin{sidewaystable}%[H]
%\resizebox{17cm}{!} {
\begin{tabular}{rrrrrrrrrrrrrrrr}
\toprule
	 & \multicolumn{5}{c}{\pmb{$k= 0.5$}} & \multicolumn{5}{c}{\pmb{$k= 0.75$}} & \multicolumn{5}{c}{\pmb{$k=0.85$}}\\
\pmb{Model} & \textbf{C} & \textbf{S} & \textbf{W} & \textbf{SI} & \textbf{MI} & \textbf{C} & \textbf{S} & \textbf{W} & \textbf{SI} & \textbf{MI} & \textbf{C} & \textbf{S} & \textbf{W} & \textbf{SI} & \textbf{MI} \\
\midrule
% & c & s & w & si & mi & c & s & w & si & mi & c & s & w & si & mi \\ 
%  \hline
	\textbf{M1} & 21.3 & 22.5 & 22.5 & 21.3 & 21.1 & 19.1 & 20.1 & 20.2 & 18.2 & 17.9 & 19.9 & 21.0 & 21.0 & 19.1 & 18.5 \\ 
  \textbf{M2} & 21.3 & 22.1 & 22.1 & 21.6 & 21.1 & 19.1 & 19.9 & 19.9 & 18.9 & 18.4 & 19.9 & 20.9 & 20.9 & 20.3 & 19.2 \\ 
  \textbf{M3} & 21.3 & 22.1 & 21.6 & 21.4 & 21.4 & 19.1 & 19.9 & 19.4 & 19.0 & 18.8 & 19.9 & 20.9 & 20.2 & 19.8 & 19.4 \\ 
  \textbf{M4} & 21.3 & 22.0 & 22.0 & 21.6 & 21.5 & 19.1 & 19.5 & 19.5 & 18.8 & 18.3 & 19.9 & 20.5 & 20.5 & 20.1 & 19.1 \\ 
   \hline
\end{tabular}
%}
\caption{$\mbox{ASE}\left(\widehat{\mathcal M}\right)$ ($\times 1000$) for complete (\textbf{C}), simplified (\textbf{S}), IPW (\textbf{W}), single imputation (\textbf{SI}) and multiple imputation (\textbf{MI}) methods in scenario 3.} 
\label{tabla:ESC3}
\end{sidewaystable}

%............................................%
\section{Illustration with real data}
\label{sec:data}
Two applications of the new methodology are presented in this section. The first example is taken from medicine and refers to the study of HIV-AIDS patients with advanced immune suppression. The second example considers marine observations, focusing on maximum surface temperature  (in ºC) at a marine buoy near the Cies Islands in Pontevedra (NW Spain). Both data examples present missing data in the response variable to be modelled.

\subsection{Example 1. Study of HIV-AIDS}
\label{subsec:data1}
\cite{zhang2022} considered this dataset in a regression setting with response missing at random. The dataset is accessible via the R package \textit{BART}, \cite{Sparapani:2021}. From an HIV trial, 522 subjects with didanosine antiretroviral regimen have been selected. After the treatments were applied, the CD4 count at 96 $\pm 5$ weeks and other variables were collected from each patient. For this HIV clinical trail, the CD4 cell counts is the target variable, decreasing as HIV progresses. To develop our analysis, conditional modes of CD4 count at 96 $\pm 5$ weeks ($\texttt{cd496}=Y$) were estimated given age variable ($\texttt{age}={X}$).

\begin{figure}[H]
\centering
\includegraphics[width=0.9\linewidth]{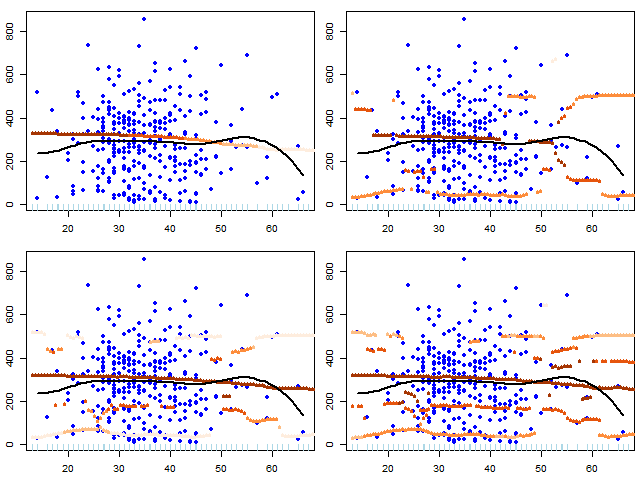}
\caption {Conditional modes with multiple imputation of $\texttt{cd496}$ given $\texttt{age}$ with modes number 1,2,3,4 (from top left corner to bottom right corner) respectively. Smooth regression estimates in black (with only complete cases), and modal regression in orange. Orange light colors are shown for low density values and orange dark colors for high density values.}
\label{tab:nmodes} 
\end{figure}

\begin{figure}[H]
\centering
\includegraphics[width=0.70\linewidth]{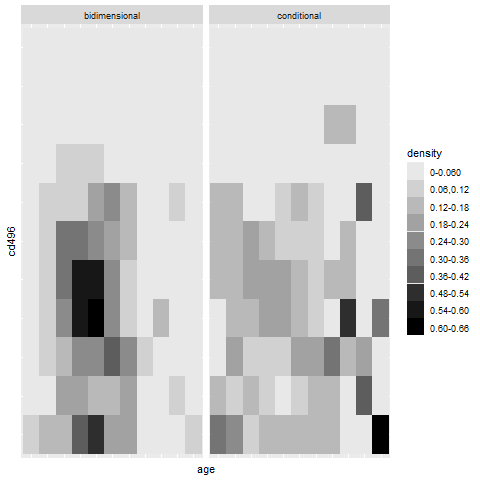}
\caption{Histogram of $(\texttt{age},\texttt{cd496})$ (left) and  conditional histogram $\texttt{cd496}$ given $\texttt{age}$ (right).}
\label{fig:conditionaldensity}
\end{figure}

A total of 211 observations are classified as missing in the response variable, representing approximately 39.7\% of the total sample. Following \cite{zhang2022}, the MAR mechanism is assumed to govern the missing data. Figure~\ref{fig:conditionaldensity} shows (left) the two-dimensional density estimate (histogram) of $(\texttt{age},\texttt{cd496})$ and (right) the conditional density estimate of $\texttt{cd496}$ given $\texttt{age}$. These two plots show different types of structures of the bidimensional and conditional densities, justifying a modal regression approach. A scatter plot of the data is presented in Table~\ref{tab:nmodes}  which shows the conditional modes estimations (with multiple imputation) versus the nonparametric regression estimation. Results of the other estimators are shown in the supplementary material. 
The conditional mode estimators are obtained by assuming one to four modes in the algorithm as illustrated in the figure from the top left to the bottom right. Upon examination of the data, the conditional mode estimation algorithms are able to discern branches that are not discernible through mean regression. 

%\begin{table}
%    \centering
%    \begin{tabular}{c|c}
%       \includegraphics[width=0.5\linewidth]{cd496-modes-start1.png}   &  \includegraphics[width=0.5\linewidth]{cd496-modes-start2.png} \\
%       \hline
%       \includegraphics[width=0.5\linewidth]{cd496-modes-start3.png} & \includegraphics[width=0.5\linewidth]{cd496-modes-start4.png} \\
%    \end{tabular}
%    \caption{Conditional modes of $\texttt{cd496}$ given $\texttt{age}$ with modes number 1,2,3,4 (from top left corner to bottom right corner) respectively.}
%    \label{tab:nmodes}
%\end{table}

%----------------------------------%
\subsection{Example 2. Maximum surface temperature}
\label{subsec:data2}
MeteoGalicia, the Galizian Meteorology Agency, provides daily data from weather stations and marine buoys covering the whole of Galicia and its neighbouring regions.
The floating devices collect data from different measurements providing: oxygenous
humidity, temperature, etc.  The aim of this study is to examine the relationship between the daily maximum surface temperature ($Y$) and the average air temperature ($X$) measured on a marine buoy.
A ocean--meteorological buoy situated in the vicinity of the Cies Islands, located within the municipality of Vigo/Spain (longitude/latitude: -8° 53.59', 42° 10.69') have been considered. The data correspond to daily measurements in 2021. The buoy did not collect data for any of the 25 variables collected by the system on 8 days of the year. Therefore, the number of observations is $357$ with a $21.3\%$ incidence of missing data for the response variable $Y$.
The data are available in the oceanographic information of \url{https://www.meteogalicia.gal/}. Due to the nature of the data, there may be a degree of data dependency. Without missing observations, \cite{Ullah2022} developed a nonlinear modal regression under stationary $\alpha-$mixing dependent samples. The study of situation of missing and dependent observations is a future  research direction. 

%The aim of this study is to examine the relationship between the daily maximum surface temperature ($Y$) and the average air temperature ($X$)\footnote{\textcolor{red}{Also collected at the buoy?}}.

\begin{figure}[H]
\centering
\includegraphics[width=0.9\linewidth]{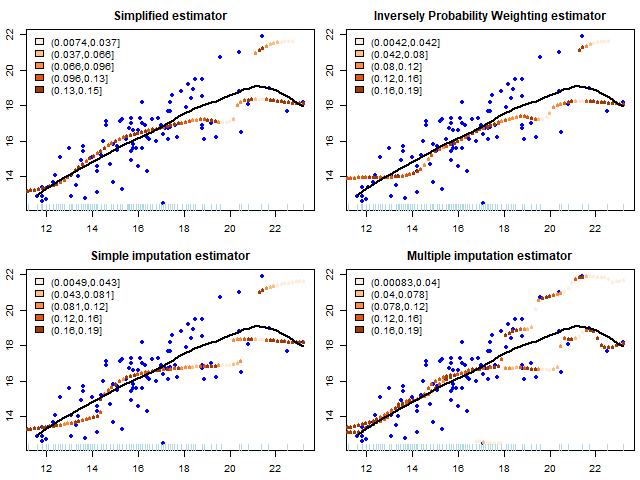}
\caption{Scatterplot of maximum surface temperature with respect to mean air temperature in 2021 at Cies marine buoy (blue solid points). Nonparametric regression estimation (black line) and different modal regression estimation (green) with 2 modes (light orange colors are shown for low density values and dark orange colors for high density values).}
\label{figura:BoiaCies}
\end{figure}

Figure~\ref{figura:BoiaCies} depicts the scatterplot of the variables and the nonparametric regression model estimation, as well as the multimode conditional estimations proposed in Section~\ref{sec:missing}. The discrepancies are statistically significant, and our proposals, particularly the $\textbf{MI}$ distinguishing branches for high values of $x$, which the mean regression identifies as potential outliers, merit further investigation. 
% Our objective is to study the relation between  the daily maximum surface temperature ($Y$) and the average air temperature ($\mathbf{X}$). The Figure \ref{figura:BoiaCies} shows the scatterplot of the variables and the estimations of the nonparametric regression model and the multimode conditional estimations proposed in the paper. The differents are significative, our proposals, specially the  $\textbf{MI}$ distingues branches for high values of x that the mean regression detect as possible outliers.  

%............................................%
\section{Concluding Remarks}
\label{sec:discussion}
This paper presents four proposals for the estimation of the conditional modes when the response variable has missing observations.  In this study, we consider a multimodal setting and we adapt the conditional mean--shift algorithm (\cite{Einbek:2006}, \cite{Chen:2016a}) considering different missing data methodologies. Specifically, we consider a simplified estimator using the  complete sample $\textbf{S}$; an Inverse probability weighting method $\textbf{IPW}$; a simple imputation for the conditional mode of the observed subsample, $\textbf{SI}$ and a novel approach, which generates multiple imputations and then combines the results of the completed samples by considering a conditional mode, namely $\textbf{MI}$. The different proposals are compared in a simulation study, where $\textbf{MI}$ is shown to outperform its competitors.

As in the case of complete data, in the unimodal scenario, if the mode coincides with the mean, the proposed estimators provide the same curve as regression on the mean. Regression on the mean has been investigated when there are missing observations in the response from different points of view by \cite{CHU1995}, \cite{Gonzalez-Manteiga2004} or \cite{Efromovich2014} among others. 

The missing scheme considered is MAR. A more complex situation where the missingness depends on the response variable $Y$, known as Missing Not At Random could be taken into account. However, in this case, it is necessary to model the missing data mechanism and generally Bayes or maximum likelihood methodologies are used. However, the multimodal  regression has not been addressed so far with these methods even in the complete data case.

Several open problems still exist. As mentioned in the previous section, the modal regression for dependent data with missing observations has not yet been studied. The literature about this topic is scarce with recent publications such as \cite{Ullah2022} or \cite{Wang2024}. For mean regression, the authors \cite{perezgonzalez} have already examined this situation. 
It would be very interesting in the future to see  the effect of dependence and missing data in modal regression. 

Second, we consider a scalar bandwidth  chosen by a data-driven method based on a complete data selector proposed by \cite{Zhou:2019}; but alternatives based on variable bandwidths, despite the computational cost, would improve the behavior of the estimators. These are unexplored lines of research, and they suggest interesting directions for future work.

\section*{Supplementary material}
The supplementary material of this paper contains codes and data needed for executing the simulation study and real data examples. This is also included in a GitHub repository \url{http://github.com/TomasCotos/NP-Modal-Reg-Missing-Y/}.

\section*{Acknowledgments}
This research has been partially supported by Project PID2020-116587GB-I00 (MICIU/AEI/ 10.13039/501100011033) and PID2020-118101GB-I00 (MCIN/ AEI /10.13039/501100011033), funded by Ministerio de Ciencia e Innovación.

\section*{Credit author statement}
Ana P\'erez-Gonz\'alez: Methodology, formal analysis, investigation, writing original--draft, software development and validation. Tom\'as R. Cotos-Y\'a\~nez: Data curation, software development and validation, visualization. Rosa M. Crujeiras: Conceptualization, supervision, writing--reviewing and editing, funding acquisition.

\bibliography{reference}% common bib file


\begin{thebibliography}{56}
\ifx \bisbn   \undefined \def \bisbn  #1{ISBN #1}\fi
\ifx \binits  \undefined \def \binits#1{#1}\fi
\ifx \bauthor  \undefined \def \bauthor#1{#1}\fi
\ifx \batitle  \undefined \def \batitle#1{#1}\fi
\ifx \bjtitle  \undefined \def \bjtitle#1{#1}\fi
\ifx \bvolume  \undefined \def \bvolume#1{\textbf{#1}}\fi
\ifx \byear  \undefined \def \byear#1{#1}\fi
\ifx \bissue  \undefined \def \bissue#1{#1}\fi
\ifx \bfpage  \undefined \def \bfpage#1{#1}\fi
\ifx \blpage  \undefined \def \blpage #1{#1}\fi
\ifx \burl  \undefined \def \burl#1{\textsf{#1}}\fi
\ifx \doiurl  \undefined \def \doiurl#1{\url{https://doi.org/#1}}\fi
\ifx \betal  \undefined \def \betal{\textit{et al.}}\fi
\ifx \binstitute  \undefined \def \binstitute#1{#1}\fi
\ifx \binstitutionaled  \undefined \def \binstitutionaled#1{#1}\fi
\ifx \bctitle  \undefined \def \bctitle#1{#1}\fi
\ifx \beditor  \undefined \def \beditor#1{#1}\fi
\ifx \bpublisher  \undefined \def \bpublisher#1{#1}\fi
\ifx \bbtitle  \undefined \def \bbtitle#1{#1}\fi
\ifx \bedition  \undefined \def \bedition#1{#1}\fi
\ifx \bseriesno  \undefined \def \bseriesno#1{#1}\fi
\ifx \blocation  \undefined \def \blocation#1{#1}\fi
\ifx \bsertitle  \undefined \def \bsertitle#1{#1}\fi
\ifx \bsnm \undefined \def \bsnm#1{#1}\fi
\ifx \bsuffix \undefined \def \bsuffix#1{#1}\fi
\ifx \bparticle \undefined \def \bparticle#1{#1}\fi
\ifx \barticle \undefined \def \barticle#1{#1}\fi
\bibcommenthead
\ifx \bconfdate \undefined \def \bconfdate #1{#1}\fi
\ifx \botherref \undefined \def \botherref #1{#1}\fi
\ifx \url \undefined \def \url#1{\textsf{#1}}\fi
\ifx \bchapter \undefined \def \bchapter#1{#1}\fi
\ifx \bbook \undefined \def \bbook#1{#1}\fi
\ifx \bcomment \undefined \def \bcomment#1{#1}\fi
\ifx \oauthor \undefined \def \oauthor#1{#1}\fi
\ifx \citeauthoryear \undefined \def \citeauthoryear#1{#1}\fi
\ifx \endbibitem  \undefined \def \endbibitem {}\fi
\ifx \bconflocation  \undefined \def \bconflocation#1{#1}\fi
\ifx \arxivurl  \undefined \def \arxivurl#1{\textsf{#1}}\fi
\csname PreBibitemsHook\endcsname

%%% 1
\bibitem[\protect\citeauthoryear{Ameijeiras-Alonso
  et~al.}{2021}]{AmeijeirasAlonso:2021}
\begin{barticle}
\bauthor{\bsnm{Ameijeiras-Alonso}, \binits{J.}},
\bauthor{\bsnm{Crujeiras}, \binits{R.M.}},
\bauthor{\bsnm{Rodriguez-Casal}, \binits{A.}}:
\batitle{multimode: An r package for mode assessment}.
\bjtitle{Journal of Statistical Software}
\bvolume{97}(\bissue{9}),
\bfpage{1}--\blpage{32}
(\byear{2021})
\end{barticle}
\endbibitem

%%% 2
\bibitem[\protect\citeauthoryear{Alonso~Pena}{2020}]{alonsopena-2020}
\begin{barticle}
\bauthor{\bsnm{Alonso~Pena}, \binits{M.}}:
\batitle{An introduction to nonparametric multimodal regression}.
\bjtitle{BEIO, Boletín de Estadística e Investigación Operativa}
\bvolume{36}(\bissue{1}),
\bfpage{5}--\blpage{23}
(\byear{2020})
\end{barticle}
\endbibitem

%%% 3
\bibitem[\protect\citeauthoryear{Alonso-Pena and
  Crujeiras}{2023}]{alonso-pena_crujeiras_2023}
\begin{barticle}
\bauthor{\bsnm{Alonso-Pena}, \binits{M.}},
\bauthor{\bsnm{Crujeiras}, \binits{R.M.}}:
\batitle{Analyzing animal escape data with circular nonparametric multimodal
  regression}.
\bjtitle{Annals of Applied Statistics}
\bvolume{17}(\bissue{1}),
\bfpage{130}--\blpage{152}
(\byear{2023})
\end{barticle}
\endbibitem

%%% 4
\bibitem[\protect\citeauthoryear{Bianco et~al.}{2011}]{bianco2011asymptotic}
\begin{barticle}
\bauthor{\bsnm{Bianco}, \binits{A.}},
\bauthor{\bsnm{Boente}, \binits{G.}},
\bauthor{\bsnm{Gonz{\'a}lez-Manteiga}, \binits{W.}},
\bauthor{\bsnm{P{\'e}rez-Gonz{\'a}lez}, \binits{A.}}:
\batitle{Asymptotic behavior of robust estimators in partially linear models
  with missing responses: the effect of estimating the missing probability on
  the simplified marginal estimators}.
\bjtitle{Test}
\bvolume{20},
\bfpage{524}--\blpage{548}
(\byear{2011})
\end{barticle}
\endbibitem

%%% 5
\bibitem[\protect\citeauthoryear{Boente et~al.}{2009}]{boente_estimation}
\begin{barticle}
\bauthor{\bsnm{Boente}, \binits{G.}},
\bauthor{\bsnm{Gonz\'alez-Manteiga}, \binits{W.}},
\bauthor{\bsnm{P\'erez-Gonz\'alez}, \binits{A.}}:
\batitle{Robust nonparametric estimation with missing data}.
\bjtitle{J. Statist. Plann. Inference}
\bvolume{139}(\bissue{2}),
\bfpage{571}--\blpage{592}
(\byear{2009})
\end{barticle}
\endbibitem

%%% 6
\bibitem[\protect\citeauthoryear{Bashtannyk and
  Hyndman}{2001}]{bashtannyk2001bandwidth}
\begin{barticle}
\bauthor{\bsnm{Bashtannyk}, \binits{D.M.}},
\bauthor{\bsnm{Hyndman}, \binits{R.J.}}:
\batitle{Bandwidth selection for kernel conditional density estimation}.
\bjtitle{Computational Statistics \& Data Analysis}
\bvolume{36}(\bissue{3}),
\bfpage{279}--\blpage{298}
(\byear{2001})
\end{barticle}
\endbibitem

%%% 7
\bibitem[\protect\citeauthoryear{Bojinov et~al.}{2019}]{Bojinov}
\begin{barticle}
\bauthor{\bsnm{Bojinov}, \binits{I.I.}},
\bauthor{\bsnm{Pillai}, \binits{N.S.}},
\bauthor{\bsnm{Rubin}, \binits{D.B.}}:
\batitle{Diagnosing missing always at random in multivariate data}.
\bjtitle{Biometrika}
\bvolume{107}(\bissue{1}),
\bfpage{246}--\blpage{253}
(\byear{2019})
\end{barticle}
\endbibitem

%%% 8
\bibitem[\protect\citeauthoryear{Chu and Cheng}{1995}]{CHU1995}
\begin{barticle}
\bauthor{\bsnm{Chu}, \binits{C.K.}},
\bauthor{\bsnm{Cheng}, \binits{P.E.}}:
\batitle{Nonparametric regression estimation with missing data}.
\bjtitle{Journal of Statistical Planning and Inference}
\bvolume{48}(\bissue{1}),
\bfpage{85}--\blpage{99}
(\byear{1995})
\end{barticle}
\endbibitem

%%% 9
\bibitem[\protect\citeauthoryear{Chen et~al.}{2016}]{Chen:2016a}
\begin{barticle}
\bauthor{\bsnm{Chen}, \binits{Y.-C.}},
\bauthor{\bsnm{Genovese}, \binits{C.R.}},
\bauthor{\bsnm{Tibshirani}, \binits{R.J.}},
\bauthor{\bsnm{Wasserman}, \binits{L.}}:
\batitle{{Nonparametric modal regression}}.
\bjtitle{The Annals of Statistics}
\bvolume{44}(\bissue{2}),
\bfpage{489}--\blpage{514}
(\byear{2016})
\end{barticle}
\endbibitem

%%% 10
\bibitem[\protect\citeauthoryear{Chacón}{2020}]{Chacon2020}
\begin{barticle}
\bauthor{\bsnm{Chacón}, \binits{J.E.}}:
\batitle{The modal age of statistics}.
\bjtitle{International Statistical Review}
\bvolume{88}(\bissue{1}),
\bfpage{122}--\blpage{141}
(\byear{2020})
\end{barticle}
\endbibitem

%%% 11
\bibitem[\protect\citeauthoryear{Cheng}{1995}]{cheng1995}
\begin{barticle}
\bauthor{\bsnm{Cheng}, \binits{Y.}}:
\batitle{Mean shift, mode seeking, and clustering}.
\bjtitle{IEEE transactions on pattern analysis and machine intelligence}
\bvolume{17}(\bissue{8}),
\bfpage{790}--\blpage{799}
(\byear{1995})
\end{barticle}
\endbibitem

%%% 12
\bibitem[\protect\citeauthoryear{Chen}{2018}]{Chen2018}
\begin{barticle}
\bauthor{\bsnm{Chen}, \binits{Y.}}:
\batitle{Modal regression using kernel density estimation: A review}.
\bjtitle{WIREs Computational Statistics}
\bvolume{10}(\bissue{4}),
\bfpage{1431}
(\byear{2018})
\end{barticle}
\endbibitem

%%% 13
\bibitem[\protect\citeauthoryear{Comaniciu and Meer}{2002}]{comaniciu2002}
\begin{barticle}
\bauthor{\bsnm{Comaniciu}, \binits{D.}},
\bauthor{\bsnm{Meer}, \binits{P.}}:
\batitle{Mean shift: A robust approach toward feature space analysis}.
\bjtitle{IEEE Transactions on pattern analysis and machine intelligence}
\bvolume{24}(\bissue{5}),
\bfpage{603}--\blpage{619}
(\byear{2002})
\end{barticle}
\endbibitem

%%% 14
\bibitem[\protect\citeauthoryear{Carpenter and
  Smuk}{2021}]{carpenter2021missing}
\begin{barticle}
\bauthor{\bsnm{Carpenter}, \binits{J.R.}},
\bauthor{\bsnm{Smuk}, \binits{M.}}:
\batitle{Missing data: A statistical framework for practice}.
\bjtitle{Biometrical Journal}
\bvolume{63}(\bissue{5}),
\bfpage{915}--\blpage{947}
(\byear{2021})
\end{barticle}
\endbibitem

%%% 15
\bibitem[\protect\citeauthoryear{Chen et~al.}{2015}]{Chen_quantile}
\begin{barticle}
\bauthor{\bsnm{Chen}, \binits{X.}},
\bauthor{\bsnm{Wan}, \binits{A.T.K.}},
\bauthor{\bsnm{Zhou}, \binits{Y.}}:
\batitle{Efficient quantile regression analysis with missing observations}.
\bjtitle{J. Amer. Statist. Assoc.}
\bvolume{110}(\bissue{510}),
\bfpage{723}--\blpage{741}
(\byear{2015})
\end{barticle}
\endbibitem

%%% 16
\bibitem[\protect\citeauthoryear{Dubnicka}{2009}]{Dubnicka-2009}
\begin{barticle}
\bauthor{\bsnm{Dubnicka}, \binits{S.R.}}:
\batitle{Kernel density estimation with missing and auxiliary variables}.
\bjtitle{Australian \& New Zealand Journal of Statistics}
\bvolume{51}(\bissue{3}),
\bfpage{247}--\blpage{270}
(\byear{2009})
\end{barticle}
\endbibitem

%%% 17
\bibitem[\protect\citeauthoryear{Efromovich}{2014}]{Efromovich2014}
\begin{barticle}
\bauthor{\bsnm{Efromovich}, \binits{S.}}:
\batitle{Nonparametric regression with missing data}.
\bjtitle{WIREs Computational Statistics}
\bvolume{6}(\bissue{4}),
\bfpage{265}--\blpage{275}
(\byear{2014})
\doiurl{10.1002/wics.1303}
\end{barticle}
\endbibitem

%%% 18
\bibitem[\protect\citeauthoryear{Enders}{2022}]{enders2022applied}
\begin{bbook}
\bauthor{\bsnm{Enders}, \binits{C.K.}}:
\bbtitle{Applied Missing Data Analysis}.
\bpublisher{Guilford Publications},
\blocation{New York}
(\byear{2022})
\end{bbook}
\endbibitem

%%% 19
\bibitem[\protect\citeauthoryear{Einbeck and Tutz}{2006}]{Einbek:2006}
\begin{barticle}
\bauthor{\bsnm{Einbeck}, \binits{J.}},
\bauthor{\bsnm{Tutz}, \binits{G.}}:
\batitle{Modelling beyond regression functions: an application of multimodal
  regression to speed--flow data}.
\bjtitle{Journal of the Royal Statistical Society: Series C (Applied
  Statistics)}
\bvolume{55}(\bissue{4}),
\bfpage{461}--\blpage{475}
(\byear{2006})
\end{barticle}
\endbibitem

%%% 20
\bibitem[\protect\citeauthoryear{Gonz\'alez-Manteiga and
  P\'erez-Gonz\'alez}{2004}]{Gonzalez-Manteiga2004}
\begin{barticle}
\bauthor{\bsnm{Gonz\'alez-Manteiga}, \binits{W.}},
\bauthor{\bsnm{P\'erez-Gonz\'alez}, \binits{A.}}:
\batitle{Nonparametric mean estimation with missing data}.
\bjtitle{Comm. Statist. Theory Methods}
\bvolume{33}(\bissue{2}),
\bfpage{277}--\blpage{303}
(\byear{2004})
\end{barticle}
\endbibitem

%%% 21
\bibitem[\protect\citeauthoryear{Gonz\'alez-Manteiga and
  P\'erez-Gonz\'alez}{2006}]{Manteiga_Test}
\begin{barticle}
\bauthor{\bsnm{Gonz\'alez-Manteiga}, \binits{W.}},
\bauthor{\bsnm{P\'erez-Gonz\'alez}, \binits{A.}}:
\batitle{Goodness-of-fit tests for linear regression models with missing
  response data}.
\bjtitle{Canad. J. Statist.}
\bvolume{34}(\bissue{1}),
\bfpage{149}--\blpage{170}
(\byear{2006})
\end{barticle}
\endbibitem

%%% 22
\bibitem[\protect\citeauthoryear{Horvitz and
  Thompson}{1952}]{HorvitzThompson1952}
\begin{barticle}
\bauthor{\bsnm{Horvitz}, \binits{D.G.}},
\bauthor{\bsnm{Thompson}, \binits{D.J.}}:
\batitle{A generalization of sampling without replacement from a finite
  universe}.
\bjtitle{Journal of the American Statistical Association}
\bvolume{47}(\bissue{260}),
\bfpage{663}--\blpage{685}
(\byear{1952})
\end{barticle}
\endbibitem

%%% 23
\bibitem[\protect\citeauthoryear{Ibrahim and
  Molenberghs}{2009}]{ibrahim2009missing}
\begin{barticle}
\bauthor{\bsnm{Ibrahim}, \binits{J.G.}},
\bauthor{\bsnm{Molenberghs}, \binits{G.}}:
\batitle{Missing data methods in longitudinal studies: a review}.
\bjtitle{Test}
\bvolume{18}(\bissue{1}),
\bfpage{1}--\blpage{43}
(\byear{2009})
\end{barticle}
\endbibitem

%%% 24
\bibitem[\protect\citeauthoryear{Jing~Lv and Guo}{2017}]{Lv2017}
\begin{barticle}
\bauthor{\bsnm{Jing~Lv}, \binits{H.Y.}},
\bauthor{\bsnm{Guo}, \binits{C.}}:
\batitle{Variable selection in partially linear additive models for modal
  regression}.
\bjtitle{Communications in Statistics - Simulation and Computation}
\bvolume{46}(\bissue{7}),
\bfpage{5646}--\blpage{5665}
(\byear{2017})
\end{barticle}
\endbibitem

%%% 25
\bibitem[\protect\citeauthoryear{Khardani}{2019}]{khardani2019}
\begin{barticle}
\bauthor{\bsnm{Khardani}, \binits{S.}}:
\batitle{A semi-parametric mode regression with censored data}.
\bjtitle{Mathematical Methods of Statistics}
\bvolume{28},
\bfpage{39}--\blpage{56}
(\byear{2019})
\end{barticle}
\endbibitem

%%% 26
\bibitem[\protect\citeauthoryear{Kleinke et~al.}{2020}]{Kleinke2020}
\begin{bbook}
\bauthor{\bsnm{Kleinke}, \binits{K.}},
\bauthor{\bsnm{Reinecke}, \binits{J.}},
\bauthor{\bsnm{Salfrán}, \binits{D.}},
\bauthor{\bsnm{Spiess}, \binits{M.}}:
\bbtitle{Applied Multiple Imputation: Advantages, Pitfalls, New Developments
  and Applications in R}.
\bpublisher{Springer},
\blocation{New York}
(\byear{2020}).
\doiurl{10.1007/978-3-030-38164-6}
\end{bbook}
\endbibitem

%%% 27
\bibitem[\protect\citeauthoryear{Little et~al.}{2024}]{Little2024}
\begin{barticle}
\bauthor{\bsnm{Little}, \binits{R.J.}},
\bauthor{\bsnm{Carpenter}, \binits{J.R.}},
\bauthor{\bsnm{Lee}, \binits{K.J.}}:
\batitle{A comparison of three popular methods for handling missing data:
  complete-case analysis, inverse probability weighting, and multiple
  imputation}.
\bjtitle{Sociol. Methods Res.}
\bvolume{53}(\bissue{3}),
\bfpage{1105}--\blpage{1135}
(\byear{2024})
\end{barticle}
\endbibitem

%%% 28
\bibitem[\protect\citeauthoryear{Lee}{1989}]{LEE1989}
\begin{barticle}
\bauthor{\bsnm{Lee}, \binits{M.}}:
\batitle{Mode regression}.
\bjtitle{Journal of Econometrics}
\bvolume{42}(\bissue{3}),
\bfpage{337}--\blpage{349}
(\byear{1989})
\end{barticle}
\endbibitem

%%% 29
\bibitem[\protect\citeauthoryear{Lee}{1993}]{Lee1993}
\begin{barticle}
\bauthor{\bsnm{Lee}, \binits{M.}}:
\batitle{Quadratic mode regression}.
\bjtitle{Journal of Econometrics}
\bvolume{57}(\bissue{1}),
\bfpage{1}--\blpage{19}
(\byear{1993})
\end{barticle}
\endbibitem

%%% 30
\bibitem[\protect\citeauthoryear{Li and Huang}{2019}]{liHuang2019}
\begin{barticle}
\bauthor{\bsnm{Li}, \binits{X.}},
\bauthor{\bsnm{Huang}, \binits{X.}}:
\batitle{Linear mode regression with covariate measurement error}.
\bjtitle{Canadian Journal of Statistics}
\bvolume{47}(\bissue{2}),
\bfpage{262}--\blpage{280}
(\byear{2019})
\end{barticle}
\endbibitem

%%% 31
\bibitem[\protect\citeauthoryear{Liu and Huang}{2024}]{LiuHuang2024}
\begin{barticle}
\bauthor{\bsnm{Liu}, \binits{Q.}},
\bauthor{\bsnm{Huang}, \binits{X.}}:
\batitle{Parametric modal regression with error in covariates}.
\bjtitle{Biometrical Journal}
\bvolume{66}(\bissue{1}),
\bfpage{2200348}
(\byear{2024})
\end{barticle}
\endbibitem

%%% 32
\bibitem[\protect\citeauthoryear{Little}{2021}]{little2021missing}
\begin{barticle}
\bauthor{\bsnm{Little}, \binits{R.J.}}:
\batitle{Missing data assumptions}.
\bjtitle{Annual Review of Statistics and Its Application}
\bvolume{8},
\bfpage{89}--\blpage{107}
(\byear{2021})
\end{barticle}
\endbibitem

%%% 33
\bibitem[\protect\citeauthoryear{Little and
  Rubin}{2019}]{little2019statistical}
\begin{bbook}
\bauthor{\bsnm{Little}, \binits{R.J.A.}},
\bauthor{\bsnm{Rubin}, \binits{D.B.}}:
\bbtitle{Statistical Analysis with Missing Data}
vol. \bseriesno{793}.
\bpublisher{John Wiley \& Sons, Ltd},
\blocation{New York}
(\byear{2019})
\end{bbook}
\endbibitem

%%% 34
\bibitem[\protect\citeauthoryear{Molenberghs and
  Kenward}{2007}]{molenberghs2007missing}
\begin{bbook}
\bauthor{\bsnm{Molenberghs}, \binits{G.}},
\bauthor{\bsnm{Kenward}, \binits{M.}}:
\bbtitle{Missing Data in Clinical Studies}.
\bpublisher{John Wiley \& Sons, Ltd},
\blocation{New York}
(\byear{2007})
\end{bbook}
\endbibitem

%%% 35
\bibitem[\protect\citeauthoryear{Pérez-González et~al.}{2009}]{perezgonzalez}
\begin{botherref}
\oauthor{\bsnm{Pérez-González}, \binits{A.}},
\oauthor{\bsnm{Vilar-Fernández}, \binits{J.M.}},
\oauthor{\bsnm{González-Manteiga}, \binits{W.}}:
Asymptotic properties of local polynomial regression with missing data and
  correlated errors.
Ann Inst Stat Math
(61),
85--109
(2009)
\end{botherref}
\endbibitem

%%% 36
\bibitem[\protect\citeauthoryear{Pigott}{2001}]{pigott2001review}
\begin{barticle}
\bauthor{\bsnm{Pigott}, \binits{T.D.}}:
\batitle{A review of methods for missing data}.
\bjtitle{Educational research and evaluation}
\bvolume{7}(\bissue{4}),
\bfpage{353}--\blpage{383}
(\byear{2001})
\end{barticle}
\endbibitem

%%% 37
\bibitem[\protect\citeauthoryear{{R Core Team}}{2024}]{R.software}
\begin{bbook}
\bauthor{\bsnm{{R Core Team}}}:
\bbtitle{R: A Language and Environment for Statistical Computing}.
\bpublisher{R Foundation for Statistical Computing},
\blocation{Vienna, Austria}
(\byear{2024}).
\bcomment{R Foundation for Statistical Computing}.
\burl{https://www.R-project.org}
\end{bbook}
\endbibitem

%%% 38
\bibitem[\protect\citeauthoryear{Robins et~al.}{1995}]{robins1995analysis}
\begin{barticle}
\bauthor{\bsnm{Robins}, \binits{J.M.}},
\bauthor{\bsnm{Rotnitzky}, \binits{A.}},
\bauthor{\bsnm{Zhao}, \binits{L.P.}}:
\batitle{Analysis of semiparametric regression models for repeated outcomes in
  the presence of missing data}.
\bjtitle{Journal of the american statistical association}
\bvolume{90}(\bissue{429}),
\bfpage{106}--\blpage{121}
(\byear{1995})
\end{barticle}
\endbibitem

%%% 39
\bibitem[\protect\citeauthoryear{Rubin}{1987}]{Rubin1987}
\begin{bbook}
\bauthor{\bsnm{Rubin}, \binits{D.}}:
\bbtitle{Multiple Imputation for Nonresponse in Surveys}.
\bpublisher{John Wiley \& Sons},
\blocation{New York}
(\byear{1987})
\end{bbook}
\endbibitem

%%% 40
\bibitem[\protect\citeauthoryear{Scott}{2015}]{scott2015multivariate}
\begin{bbook}
\bauthor{\bsnm{Scott}, \binits{D.W.}}:
\bbtitle{Multivariate Density Estimation: Theory, Practice, and Visualization}.
\bpublisher{John Wiley \& Sons, Ltd},
\blocation{New York}
(\byear{2015})
\end{bbook}
\endbibitem

%%% 41
\bibitem[\protect\citeauthoryear{Seaman et~al.}{2013}]{Seaman2013}
\begin{barticle}
\bauthor{\bsnm{Seaman}, \binits{S.}},
\bauthor{\bsnm{Galati}, \binits{J.}},
\bauthor{\bsnm{Jackson}, \binits{D.}},
\bauthor{\bsnm{Carlin}, \binits{J.}}:
\batitle{What is meant by "missing at random"?}
\bjtitle{Statistical Science}
\bvolume{28}(\bissue{2}),
\bfpage{257}--\blpage{268}
(\byear{2013})
\end{barticle}
\endbibitem

%%% 42
\bibitem[\protect\citeauthoryear{Sparapani et~al.}{2021}]{Sparapani:2021}
\begin{barticle}
\bauthor{\bsnm{Sparapani}, \binits{R.}},
\bauthor{\bsnm{Spanbauer}, \binits{C.}},
\bauthor{\bsnm{McCulloch}, \binits{R.}}:
\batitle{Nonparametric machine learning and efficient computation with
  {B}ayesian additive regression trees: The {BART} {R} package}.
\bjtitle{Journal of Statistical Software}
\bvolume{97}(\bissue{1}),
\bfpage{1}--\blpage{66}
(\byear{2021})
\end{barticle}
\endbibitem

%%% 43
\bibitem[\protect\citeauthoryear{Seaman and White}{2011}]{Seaman2011}
\begin{barticle}
\bauthor{\bsnm{Seaman}, \binits{S.R.}},
\bauthor{\bsnm{White}, \binits{I.R.}}:
\batitle{Review of inverse probability weighting for dealing with missing
  data}.
\bjtitle{Statistical Methods in Medical Research}
\bvolume{22}(\bissue{3}),
\bfpage{278}--\blpage{295}
(\byear{2011})
\doiurl{10.1177/0962280210395740}
\end{barticle}
\endbibitem

%%% 44
\bibitem[\protect\citeauthoryear{Shao and Wang}{2016}]{Shao2016}
\begin{barticle}
\bauthor{\bsnm{Shao}, \binits{J.}},
\bauthor{\bsnm{Wang}, \binits{L.}}:
\batitle{{Semiparametric inverse propensity weighting for nonignorable missing
  data}}.
\bjtitle{Biometrika}
\bvolume{103}(\bissue{1}),
\bfpage{175}--\blpage{187}
(\byear{2016})
\end{barticle}
\endbibitem

%%% 45
\bibitem[\protect\citeauthoryear{Titterington and Mill}{1983}]{Titterington}
\begin{barticle}
\bauthor{\bsnm{Titterington}, \binits{D.M.}},
\bauthor{\bsnm{Mill}, \binits{G.M.}}:
\batitle{Kernel-based density estimates from incomplete data}.
\bjtitle{Journal of the Royal Statistical Society: Series B (Methodological)}
\bvolume{45}(\bissue{2}),
\bfpage{258}--\blpage{266}
(\byear{1983})
\end{barticle}
\endbibitem

%%% 46
\bibitem[\protect\citeauthoryear{Ullah et~al.}{2022}]{Ullah2022}
\begin{barticle}
\bauthor{\bsnm{Ullah}, \binits{A.}},
\bauthor{\bsnm{Wang}, \binits{T.}},
\bauthor{\bsnm{Yao}, \binits{W.}}:
\batitle{Nonlinear modal regression for dependent data with application for
  predicting covid-19}.
\bjtitle{Journal of the Royal Statistical Society Series A: Statistics in
  Society}
\bvolume{185}(\bissue{3}),
\bfpage{1424}--\blpage{1453}
(\byear{2022})
\end{barticle}
\endbibitem

%%% 47
\bibitem[\protect\citeauthoryear{Ullah et~al.}{2023}]{ULLAH2023}
\begin{barticle}
\bauthor{\bsnm{Ullah}, \binits{A.}},
\bauthor{\bsnm{Wang}, \binits{T.}},
\bauthor{\bsnm{Yao}, \binits{W.}}:
\batitle{Semiparametric partially linear varying coefficient modal regression}.
\bjtitle{Journal of Econometrics}
\bvolume{235}(\bissue{2}),
\bfpage{1001}--\blpage{1026}
(\byear{2023})
\end{barticle}
\endbibitem

%%% 48
\bibitem[\protect\citeauthoryear{van Buuren}{2018}]{van2018flexible}
\begin{bbook}
\bauthor{\bsnm{Buuren}, \binits{S.}}:
\bbtitle{Flexible Imputation of Missing Data, Second Edition}.
\bpublisher{Chapman and Hall/CRC},
\blocation{New York}
(\byear{2018}).
\doiurl{10.1201/9780429492259}
\end{bbook}
\endbibitem

%%% 49
\bibitem[\protect\citeauthoryear{Wang}{2024}]{Wang2024}
\begin{barticle}
\bauthor{\bsnm{Wang}, \binits{T.}}:
\batitle{Nonlinear kernel mode-based regression for dependent data}.
\bjtitle{Journal of Time Series Analysis}
\bvolume{45}(\bissue{2}),
\bfpage{189}--\blpage{213}
(\byear{2024})
\end{barticle}
\endbibitem

%%% 50
\bibitem[\protect\citeauthoryear{Wang et~al.}{2017}]{wang2017regularized}
\begin{barticle}
\bauthor{\bsnm{Wang}, \binits{X.}},
\bauthor{\bsnm{Cheoqiann}, \binits{H.}},
\bauthor{\bsnm{Cai}, \binits{W.}},
\bauthor{\bsnm{Shen}, \binits{D.}},
\bauthor{\bsnm{Huang}, \binits{H.}}:
\batitle{Regularized modal regression with applications in cognitive impairment
  prediction}.
\bjtitle{Advances in neural information processing systems}
\bvolume{30},
\bfpage{1448}--\blpage{1458}
(\byear{2017})
\end{barticle}
\endbibitem

%%% 51
\bibitem[\protect\citeauthoryear{Wang et~al.}{1998}]{wang1998local}
\begin{botherref}
\oauthor{\bsnm{Wang}, \binits{C.}},
\oauthor{\bsnm{Wang}, \binits{S.}},
\oauthor{\bsnm{Gutierrez}, \binits{R.G.}},
\oauthor{\bsnm{Carroll}, \binits{R.}}:
Local linear regression for generalized linear models with missing data.
Annals of Statistics,
1028--1050
(1998)
\end{botherref}
\endbibitem

%%% 52
\bibitem[\protect\citeauthoryear{Weixin~Yao and Li}{2012}]{Yao2012}
\begin{barticle}
\bauthor{\bsnm{Weixin~Yao}, \binits{B.G.L.}},
\bauthor{\bsnm{Li}, \binits{R.}}:
\batitle{Local modal regression}.
\bjtitle{Journal of Nonparametric Statistics}
\bvolume{24}(\bissue{3}),
\bfpage{647}--\blpage{663}
(\byear{2012})
\end{barticle}
\endbibitem

%%% 53
\bibitem[\protect\citeauthoryear{Yao and Li}{2014}]{YaoLi2014}
\begin{barticle}
\bauthor{\bsnm{Yao}, \binits{W.}},
\bauthor{\bsnm{Li}, \binits{L.}}:
\batitle{A new regression model: Modal linear regression}.
\bjtitle{Scandinavian Journal of Statistics}
\bvolume{41}(\bissue{3}),
\bfpage{656}--\blpage{671}
(\byear{2014})
\end{barticle}
\endbibitem

%%% 54
\bibitem[\protect\citeauthoryear{Zhou and Huang}{2016}]{zhou2016}
\begin{botherref}
\oauthor{\bsnm{Zhou}, \binits{H.}},
\oauthor{\bsnm{Huang}, \binits{X.}}:
Nonparametric modal regression in the presence of measurement error
(2016)
\end{botherref}
\endbibitem

%%% 55
\bibitem[\protect\citeauthoryear{Zhou and Huang}{2019}]{Zhou:2019}
\begin{barticle}
\bauthor{\bsnm{Zhou}, \binits{H.}},
\bauthor{\bsnm{Huang}, \binits{X.}}:
\batitle{Bandwidth selection for nonparametric modal regression}.
\bjtitle{Communications in Statistics - Simulation and Computation}
\bvolume{48}(\bissue{4}),
\bfpage{968}--\blpage{984}
(\byear{2019})
\end{barticle}
\endbibitem

%%% 56
\bibitem[\protect\citeauthoryear{Zhang and Wang}{2022}]{zhang2022}
\begin{barticle}
\bauthor{\bsnm{Zhang}, \binits{T.}},
\bauthor{\bsnm{Wang}, \binits{L.}}:
\batitle{Smoothed partially linear quantile regression with nonignorable
  missing response}.
\bjtitle{Journal of the Korean Statistical Society}
\bvolume{51}(\bissue{2}),
\bfpage{441}--\blpage{479}
(\byear{2022})
\end{barticle}
\endbibitem

\end{thebibliography}

\end{document}